\documentclass[aps,prx,twocolumn,superscriptaddress,noshowpacs]{revtex4-2}

\usepackage{times}

\usepackage{graphicx}
\usepackage{soul,color}
\usepackage{float}
\usepackage{bm}
\usepackage{braket}
\usepackage{placeins}
\usepackage{amsmath,amssymb}
\usepackage{comment}
\usepackage[dvipsnames]{xcolor}
\usepackage{url}
\usepackage[normalem]{ulem}
\usepackage{bm,nicefrac,xfrac}
\usepackage{physics}
\usepackage{xcolor}
\usepackage[normalem]{ulem}

\usepackage{xcolor}

\usepackage[normalem]{ulem}

\begin{document}

\title{Quantum critical behavior of cuprate superconductors observed by inelastic X-ray scattering}

\author{H. Y. Huang}
\affiliation{National Synchrotron Radiation Research Center, Hsinchu 300092, Taiwan}
\author{C. Y. Mou}
\affiliation{Center for Quantum Science and Technology and Department of Physics, National Tsing Hua University, Hsinchu 300044, Taiwan}

\author{A. Singh}
\affiliation{National Synchrotron Radiation Research Center, Hsinchu 300092, Taiwan}

\author{J. S. Su}
\author{J. Okamoto}
\affiliation{National Synchrotron Radiation Research Center, Hsinchu 300092, Taiwan}

\author{S. Komiya}
\affiliation{Central Research Institute of Electric Power Industry, Yokosuka, Kanagawa, 240-0196, Japan}

\author{C. T. Chen}
\affiliation{National Synchrotron Radiation Research Center, Hsinchu 300092, Taiwan}

\author{T. K. Lee}
\affiliation{Institute of Physics, Academia Sinica, Taipei 115201, Taiwan}
\affiliation{Department of Physics, National Tsing Hua University, Hsinchu 300044, Taiwan}

\author{A. Fujimori}
\affiliation{National Synchrotron Radiation Research Center, Hsinchu 300092, Taiwan}
\affiliation{Center for Quantum Science and Technology and Department of Physics, National Tsing Hua University, Hsinchu 300044, Taiwan}
\affiliation{Department of Physics, University of Tokyo, Bunkyo-Ku, Tokyo 113-0033, Japan}

\author{D. J. Huang}
\altaffiliation [email: ] {\emph{djhuang@nsrrc.org.tw}} 
\affiliation{National Synchrotron Radiation Research Center, Hsinchu 300092, Taiwan}
\affiliation{Department of Physics, National Tsing Hua University, Hsinchu 300044, Taiwan}
\affiliation{Department of Electrophysics, National Yang Ming Chiao Tung University, Hsinchu 300093, Taiwan}

%\cite{sachdev2011, CapraraPRB2017, proust2019,Cooper2009,Michon2019, LegrosNatPhys2019}

\begin{abstract}
Progress toward a complete understanding of cuprate superconductors has been hindered by their intricate phase diagram, potentially linked to a quantum critical point (QCP). However, conclusive evidence for the QCP is lacking, as the presumed QCP is buried under the superconducting dome, disguising its presence. Here, we use high-resolution resonant inelastic X-ray scattering to examine the dynamical charge-charge correlation in La$_{2-x}$Sr$_x$CuO$_4$ and uncover the quantum critical scaling, a key feature required for a QCP.  Specifically, we observed that the inverse correlation lengths for various dopings and temperatures collapsed onto a universal scaling curve, yielding a critical exponent $\nu$ of $0.74 \pm 0.08$. The non-negativity of this exponent confirms the presence of a QCP. Remarkably, the value of $\nu$ suggests that while the QCP is manifested through the charge-density wave, other orders also participate, such that the QCP appears to belong to the universality class characterized by the O(4) symmetry, reminiscent of the microscopic SO(4) symmetry in the Hubbard model at half-filling. 
Further analysis indicates that the QCP is highly dissipative with a short quasi-particle lifetime, reflecting the intertwined quantum fluctuations due to its being buried inside the superconducting state.

\end{abstract}

\date{\today}

%\flushbottom
\maketitle

\thispagestyle{empty}
%%%%%%%%%%%%%%%%%%%%%%

\begin{figure}[t!]
\centering
\includegraphics[width=0.95 \columnwidth]{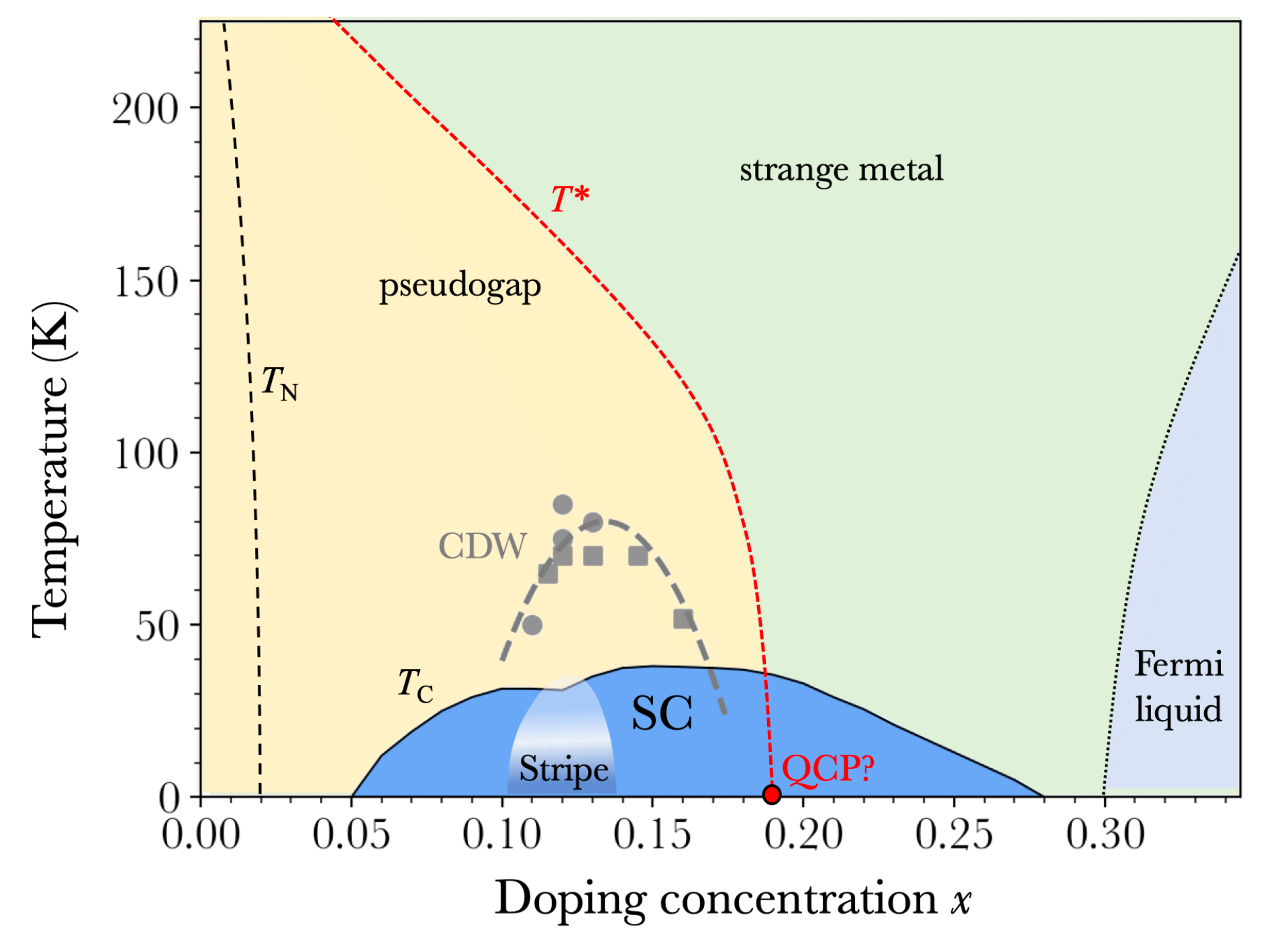}
\caption{Phase diagram of La$_{2-x}$Sr$_x$CuO$_4$ plotted in the plane of temperature versus hole doping level $x$ and its anomalous quantum criticality. The black solid curve, denoted by $T_{_{\rm C}}$, and the red dashed curve, denoted by $T^*$, mark the SC transition temperature and the crossover temperature of the pseudogap phase, respectively. The black dashed line marked by $T_{\rm N}$ indicates the N\'eel temperature. The red circle denotes the putative QCP. The gray circles and squares indicate the onset temperature of the CDW, with a dashed curve providing a visual guide \cite{Croft2014, WenNatComm2019, Choiniere2018}. Reports have indicated varying onset curves \cite{Miao2021,LinPRL2020}. Additionally, studies have observed different CDW orders (the CDW stripe orders) exhibiting different temperature dependencies across $T_c$. The CDW stripe order is in the low temperature regime with doping $x$ between 0.10 and 0.13 (see Refs.~\onlinecite{WenNatComm2019} and \onlinecite{Croft2014}). In the present work, we shall focus on the CDW order that is most often cited and is beyond the CDW stripe order. Note that different kinds of CDW order also appear in the stripe order region \cite{WenNatComm2019}. The strange-metal phase is a region in which the coefficient of the $T$-linear term in resistivity is non-vanishing and much larger than that of the quadratic term \cite{Cooper2009,ayres2021}. LSCO exhibits anomalous quantum criticality: its strange-metal phase extends to the overdoped region of $x=$~0.22.} 
\label{phase_diagram}
\end{figure}

The physics of cuprate superconductors has long held a position of paramount importance in condensed matter physics, yet it continues to be in an enigma \cite{KeimerNature2015}. Despite advances in our understanding, the underlying mechanisms driving cuprate superconductivity (SC) remain shrouded in mystery. Central to the  physics of cuprate  superconductors is their remarkably rich and complex phase diagram, plotted in the plane of temperature versus doping concentration. Figure~\ref{phase_diagram} shows the phase diagram of La$_{2-x}$Sr$_x$CuO$_4$  (LSCO), a prototypical single-layer hole-doped superconducting cuprate. When doped with holes, cuprates at low temperatures can be tuned from a Mott insulating phase to a superconducting phase, and then to a Fermi liquid phase. This landscape of phases and transitions unveils the fascinating physical properties of cuprates, ranging from SC to charge order and beyond.  

In the underdoped regime of hole-doped cuprates above the SC transition temperature $T_{_{\rm C}}$, various physical quantities show an enigmatic electronic excitation gap called a pseudogap  \cite{KeimerNature2015}. Such a pseudogap phase forms below a crossover temperature $T^*$, which decreases monotonically and ends at the critical doping $x_{\rm c}$ when the doping is increased. The origins of the pseudogap remain the subjects of intense study, adding to the mystique of cuprate physics. Additionally, several symmetry-breaking orders, such as charge-density waves (CDW)~\cite{GhiringhelliScience2012,Croft2014,FradkinRMP2015,WenNatComm2019,ArpaiaScience2019,FradkinRMP2015,LeeNatPhys2021,HYHuang2021,arpaia2023}, pair-density waves (PDW) \cite{FradkinRMP2015,TuSciRep2019,Agterberg2020,Du2020,Lee2023PDW}, and nematic order~\cite{Sato2017,Ishida2020} have been discovered in the cuprates;  their mechanism and competition with SC remain the subjects of debate. 

%\vspace{1cm}
%\subsection*{ Quantum criticality shaping perspectives on cuprate superconductors}

One of the fundamentally significant and novel properties of cuprate superconductors is their defiance of the conventional Fermi-liquid theory.  Anomalous thermodynamic \cite{Michon2019} and transport properties \cite{Cooper2009,LegrosNatPhys2019} near the critical doping have been observed in cuprates, indicating their non-Fermi liquid behavior. For example, as temperature $T$ approaches the absolute zero, the electrical resistivity of cuprates displays a $T$-linear relationship, deviating from the $T^2$ dependence characteristic of Fermi liquids \cite{Giraldo-Gallo2018}. This distinctive behavior is understood to arise as the scattering rate of charge carriers reaches the Planckian limit above the quantum critical point (QCP), irrespective of the underlying Fermi surface topology.

Upon tuning a non-thermal parameter through a critical value, quantum phase transitions occur at the absolute zero of temperature. The presence of the QCP in a cuprate holds the key to understanding many profound phenomena related to its SC \cite{sachdev2010,sachdev2011,CapraraPRB2017,proust2019,Michon2019,Cooper2009,LegrosNatPhys2019,Giraldo-Gallo2018}. It provides a more comprehensive and holistic perspective on the complex behaviors of cuprates and offers insights into their underlying physics. The concept of quantum criticality has the potential to reshape our understanding of the phase diagram of cuprate superconductors and to explain the non-Fermi liquid behavior observed in them. Figure~\ref{phase_diagram} also illustrates that the ground state of LSCO is a superconducting state for doping $x$ between 0.05 and 0.28 \cite{Choiniere2018, WenNatComm2019}. 
Interestingly, LSCO exhibits anomalous quantum criticality; its strange-metal phase at $T$ = $0$ extends to the overdoped region of $x=0.3$ \cite{Cooper2009,ayres2021}. Quantum fluctuations near a QCP can give rise to anomalous behaviors in transport, thermodynamics, and other properties. 
However, numerous experimental results have raised skepticism regarding the ideal QCP scenario  \cite{Hussey2018,Tallon2020,ayres2021}. Obtaining evidence for the QCP associated with the CDW in the superconducting phase of cuprates is challenging. Further investigation into the evolution of CDW as the QCP is approached is of importance in enhancing our understanding of CDW dynamics. 

In this Article, we present results of high-resolution resonant inelastic X-ray scattering (RIXS) of LSCO at various hole concentrations and temperatures to unravel its intriguing quantum fluctuations associated with CDW that exhibits relaxation characteristics. Through temperature- and doping-dependent measurements, we found that when the QCP is approached, the interaction of CDW with SC gives rise to unusual CDW dynamics. In this dynamics, both the relaxation rate and the correlation length of the CDW increase,  while the lifetime of CDW quasiparticles near the QCP decreases. 
We observed a key feature indicative of a QCP: the correlation length of CDW dynamics exhibits a critical scaling with an exponent $\nu$ of $0.74 \pm 0.08$ and diverges as the QCP is approached.

The dynamical structure factor $S({\bf q}, \omega)$ of X-ray scattering probes the space and time Fourier transform of the charge density-density correlation function.  Here ${\bf q}$ and ${\omega}$ are the momentum and the energy transferred to charge excitations, respectively. RIXS measures the dynamical charge fluctuations modulated by the effects of RIXS matrix elements, X-ray polarization, and orbital characters \cite{Ament11,ArpaiaScience2019,LeeNatPhys2021,HYHuang2021,arpaia2023}. In the absence of superconducting pairing, this technique investigates static charge distribution and dynamical charge fluctuations, corresponding to charge susceptibilities denoted by $\chi_0 ({\bf q}, {\omega}=0)$ and $\chi({\bf q}, {\omega})$, respectively.  
Recently, RIXS has been proved to be an effective spectroscopic tool for the exploration of charge-density fluctuations (CDF) in cuprates \cite{ArpaiaScience2019, LeeNatPhys2021, HYHuang2021,arpaia2023}.  When the superconducting order $\Delta$ is in presence, RIXS can also probe the pairing-density wave (PDW). In this case, the elastic peak at ${\bf Q}$ results from the mixture of the CDW order $\rho_{\bf Q}$ and the PDW order $\Delta_{\bf Q}$ with the structure factor $S({\bf Q}, \omega=0)$ given by a linear combination of $|\Delta^* \Delta_{\bf Q}|^2$ and $|\rho_{\bf Q}|^2$ \cite{Lee2023PDW}.
In particular, %it is found that 
CDF is strongest at the doping level where the superfluid density and the superconducting critical current are highest, reflecting the signature of an anomalous QCP \cite{arpaia2023}. However, definite proof of the existence of QCP is not established.

\begin{figure*}[ht!]
\centering
\includegraphics[width=2.1 \columnwidth]{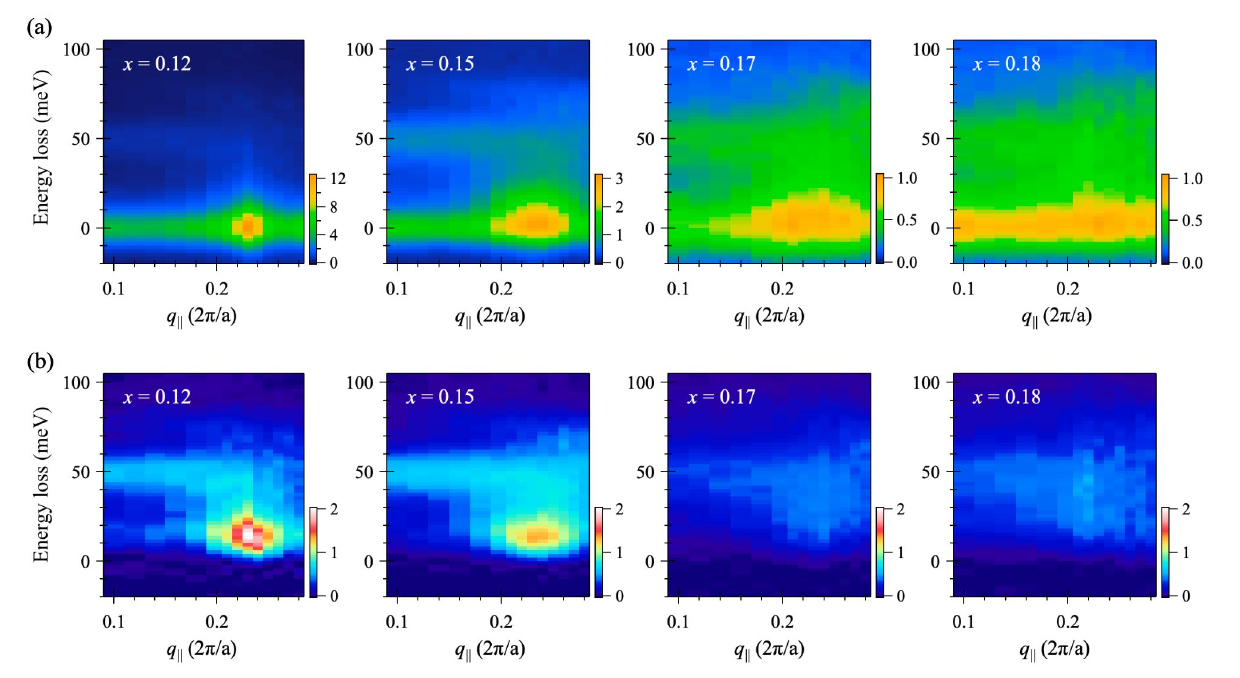}
\caption{Doping-dependent O $K$-edge RIXS of La$_{2-x}$Sr$_x$CuO$_4$ with $x = 0.12,~0.15,~0.17,~\& ~0.18$. RIXS spectra were recorded with $\sigma$-polarized incident X-ray tuned to the Zhang-Rice singlet hole peak at 24~K. The momentum transfer is \textbf{q}~$=(q_{\|}, 0, L)$ with $L$ varying between 0.47 and 1.03 in reciprocal lattice units.  (a), RIXS intensity distribution maps in the plane of energy loss vs. in-plane momentum transfer \textbf{q}$_\|$ along $(\pi, 0)$. (b), RIXS intensity distribution maps after the subtraction of elastic scattering. The raw RIXS data for each momentum scan are plotted in Figs. S2-S5 of the Supplementary Information.}
\label{rixs_map}
\end{figure*}

In a classical phase transition, anomalies in 
transport coefficients and relaxation rates occur close to a critical point, exhibiting dynamical critical phenomena \cite{Hohenberg1977}. Likewise  
we anticipate similar behaviors to occur in the dynamics of a QCP. To unravel the dynamics of CDW fluctuations in the vicinity of  QCP, we measured the high-resolution O $K$-edge RIXS of LSCO with hole concentrations approaching the critical doping at 24~K.  Figure \ref{rixs_map}(a) shows RIXS intensity in the plane of energy loss vs. in-plane momentum transfer \textbf{q}$_{\|} = (q_{\|}, 0)$.  With \textbf{q}$_{\|}$ varied along the anti-nodal direction, we observed static CDW of LSCO with wave vector ${\bf Q} = (0. 235, 0)$, expressed in units of ${2\pi}/{a}$ throughout the paper. This CDW appears incommensurate due to defects known as discommensurations in the commensurate CDW state \cite{TuSciRep2019, HYHuang2021}. We found that $|{\bf Q}|$ did not significantly depend on doping $x$, which is consistent with the saturation of the incommensurability of spin and charge order \cite{Yamada1998}.  The temperature-dependent CDW intensity and  the half width of the momentum scan, which reflects the correlation length, show that the CDW are slightly suppressed in the superconducting phase as the temperature is decreased across $T_{_{\rm C}}$ \cite{Croft2014,WenNatComm2019}, indicating the competition of CDW with SC. The CDW state and the superconducting state near $T=0$ are intertwined below the QCP. Because of small energy difference, switching back and forth between CDW and the superconducting ground states is energetically likely, resulting in quantum fluctuations.

\begin{figure*}[t!]
\centering
\includegraphics[width=2.1 \columnwidth]{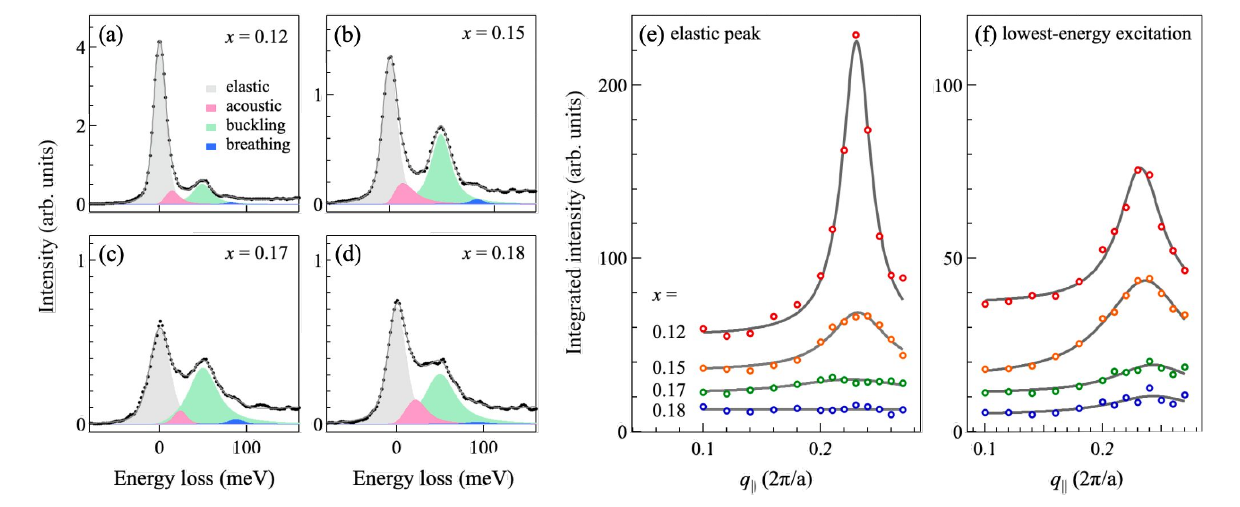}
\caption{Curve-fitting analysis of O $K$-edge RIXS of La$_{2-x}$Sr$_x$CuO$_4$ with $x=0.12, 0.15, 0.17,~\&~0.18$. (a)-(d), RIXS spectra and their spectral components after curve fitting for $q_{\|} = 0.12$. The RIXS spectra were fitted to four components with a linear background: elastic scattering, acoustic phonons, the mix of buckling phonon and apical oxygen phonon, and half-breathing (bond stretching) phonons, represented in gray, pink, green, and blue shades, respectively. Each phonon component was fitted to a spectral function of damped harmonic oscillator. RIXS data are plotted as black circles, and the fitted curve is given as a gray line. Detailed information about the curve fitting is provided in the Supplementary Information. (e) $\&$ (f), Integrated spectral weights of the fitted elastic scattering and the lowest-energy RIXS excitation, i.e., acoustic phonon for  $q_{\|}$ away from $|\textbf{Q}|$ and CDW fluctuations for $q_{\|}=|\textbf{Q}|$, as a function of  $q_{\|}$ for various doping levels. }
\label{curve_fitting}
\end{figure*} 

At low temperatures, the quantum nature of charge fluctuation is important because the thermal energy is less than the excitation energy $\omega$, which is dominated by the amplitude fluctuation of the charge density. Previous RIXS results showed the spectral signature of the CDW quantum fluctuations in LSCO \cite{HYHuang2021}. Therefore, we separated the spectral contributions from $\chi_0 ({\bf q}, {\omega}=0)$ and $\chi({\bf q}, {\omega})$. Figure \ref{rixs_map}(b) displays the RIXS intensity maps after subtraction of the contribution of elastic scattering. 
One hallmark of a QCP is the expected divergence in susceptibility to external perturbations at a phase transition. However, the measured RIXS intensity of CDW fluctuations does not increase near the QCP (see Fig.~S1 in the Supplementary Information), contrary to the expectations that quantum fluctuations would enhance scattering intensity at a QCP.  At first glance, our observation seems to support the absence of the QCP. In the following, we will show that this is disguised by intertwining of CDW and superconducting orders. Through quantitative analysis, we show that the intertwining leads to enhanced CDW relaxation, causing the resonance peak to broaden near the QCP.

%\vspace{1cm}
%\subsection*{ Quantum fluctuation of charge-density waves}

To perform a quantitative analysis of the RIXS spectral features associated with the dynamical CDW, we initially employed curve fitting to extract the spectral components of RIXS resulting from phonon excitations mediated by electron-phonon coupling. Figures~\ref{curve_fitting}(a)-\ref{curve_fitting}(d) plot the fitted spectra of $q_{\|}=0.12 $, away from $|\textbf{Q}|$. In addition to elastic scattering, the dominant spectral weights of the measured RIXS comprise several phonon modes including acoustic, buckling, apical oxygen, and breathing phonons. The phonon energies from our fitting are consistent with those of neutron scattering, X-ray scattering, and other measurements \cite{mcqueeney1999,giustino2008,sugai2013,ParkPRB2014}, revealing the phonon softening induced by charge correlations \cite{HYHuang2021}. Figures 3(e) and 3(f) depict the intensity of elastic scattering and the lowest-energy RIXS excitation as a function of  $q_{\|}$ for various doping levels. These $q_{\|}$-dependent plots indicate that the lowest-energy excitations show a peak feature at $|\textbf{Q}|$, highlighting the change in the RIXS excitation from being dominated by acoustic phonons to being dominated by CDW fluctuations as $q_{\|}$ approaches $|\textbf{Q}|$. Additionally, the static CDW vanishes for doping beyond $x=0.17$, while the dynamical CDW persists near $|\textbf{Q}|$, suggesting the evolution of the static order into the continuum of quantum fluctuations \cite{LeeNatPhys2021}.

To describe such CDW quantum fluctuations, we resorted to a phenomenological form of charge susceptibility based on the 
Ginzburg-Landau approach \cite{sachdev2011,ArpaiaScience2019,HYHuang2021,arpaia2023}.  When the system is far from the QCP, the lifetime of charge excitation, which is inversely proportional to the relaxation rate $\Gamma$, is long, leading to the expectation that the relativistic form of the charge susceptibility $\chi_{_{\rm CDW}}$ in the $q$-$\omega$ space is retained. Considering the Taylor expansion of $\chi_{_{\rm CDW}}$ in terms of $\omega$ and $q$ separately, one can exploit the Ginzburg-Landau Hamiltonian to derive the charge susceptibility. See the Supplementary Information for details. For the CDW with order parameter $\rho_{_{\textbf{Q}}}$, its charge susceptibility takes the following form \cite{sachdev2011,HYHuang2021}
\begin{equation} \label{chi}
\chi_{_{\rm CDW}}({\bf q}_{\|}, {\omega})=  \frac{1}{m^2+c^{2}q^{2}-(\omega+i\Gamma)^2},
\end{equation}
where $m$, referring to the mass term in the propagator, denotes the characteristic energy of the dynamical CDW, $q$ represents $\lvert {\rm \textbf{q}}_{\|} - \textbf{Q}\rvert$, $c$ is a parameter that characterizes the speed of excitations of CDW, and $\Gamma$ characterizes the inverse lifetime for the excitations. In the absence of fluctuations of order parameter in space and time, $m$ is uniform in space and time. In this situation, the pole of $\chi_{_{\rm CDW}}$ occurs at $\omega = \sqrt{m^2+c^{2}q^{2}}$ reflecting the excitation energy \cite{sachdev2011}, and $m$ can be identified as the gap of CDW amplitude excitations, i.e., $m = |\rho_{_{\textbf{Q}}}|$. These bosonic excitations are named quasiparticles in the following. When the CDW order fluctuates in space and time, its correlation decays over correlation length $\xi$, which depends on doping $x$ and temperature $T$. One expects that a pole of $q$ occurring at $2{\pi}i/\xi$ in Eq.~(1) gives rise to the decay of correlation length. By including an additional term due to $\xi$, $m^2$ can be expressed as $|\rho_{_{\textbf{Q}}}|^2 + 4\pi^{2}c^{2}\xi^{-2}$. As $x$ and $T$ approach the QCP, one expects $|\rho_{_{\textbf{Q}}}| \propto \xi^{-z}$ \cite{sachdev2011}, with $z$ being the dynamical exponent. For the QCP that
involves with the CDW order, $z = 1$ and $|\rho_{_{\textbf{Q}}}|$ is proportional to $\xi^{-1}$. Hence, $m$ is inversely proportional to $\xi$. 

In general, $m=m(x,T)$ is a function of doping $x$ and temperature $T$. The critical scaling \cite{Stanley1999} implies that 
\begin{equation} \label{m}
m(x,T)=T^{\nu} F(|x-x_c|/T^{\alpha}), 
\end{equation}
where $x_c$ is the doping level at the QCP, the exponent $\nu$ is the scaling dimension of $m$, $\alpha$ characterizes the potential anisotropy between doping and temperature, and $F(y)$ is a universal scaling function.  To find the asymptotic form of $m(x,T)$ near zero temperature and the critical doping, %$x~{\sim}~x_c$ and $T~{\sim}~0$, 
we note that $m(x,T) = m(x_c,T) + \delta m(x,T)$, where $\delta m(x,T)$ is the deviation of  $m(x,T)$ from  $m(x_c,T)$, i.e., $\delta m(x,T)=m(x,T) - m(x_c,T)$.  Since for temperature near zero %$T~\sim~0$, 
$m(x,T)~\sim~m(x,0)$ and $m(x_c,T)~\sim~m(x_c,0)=0$, we find that $\delta m(x,T)~\sim~m(x,0)$, and hence $m(x,T)$ obeys additivity in its parameters $x$ and $T$, i.e., $m(x,T)~\sim~m(x_c,T)+m(x,0)$. Indeed, as shown in the Supplementary Information and Ref.~\onlinecite{sachdev2011}, when $x$ and $T$ are close to the critical point, the correlation length $\xi$ can be expressed as $\xi^{-1} = \frac{2}{\pi}(\xi_{x}^{-1}+\xi_{T}^{-1})$. Here $\xi_x~\sim~|x-x_c|^{-\nu}$ is the correlation lenth of $\rho_{_{\textbf{Q}}}$ projected at zero temperature for $x~\sim~x_c$, while $\xi_T~\sim~T^{-\nu}$ is the correlation length projected along the temperature axis near zero temperature. Therefore, $m$ takes the asymptotic form of ${m \approx A|x-x_{\rm c}|^{\nu} + B T^{\nu}}$ with $A$ and $B$ being constants. This implies that $\alpha=1$ and $F(y) = A y^{\nu} +B $.

The pole of $\chi_{_{\rm CDW}}$ in the limit of $\Gamma \rightarrow 0$ has a simple meaning. It gives rise to  quasi-particle energy with $\omega = \sqrt{m^2+c^{2}q^{2}}$, reflecting the relativistic dynamics such that $\omega$ is proportional to $q$ when $m=0$ ; this is similar to antiferromagnetic magnon. 
When the doping $x$ approaches the critical doping, quantum critical fluctuations activate competing orders temporarily so that CDW order can transform into other orders through scattering (which is termed as intertwining). Thus one expects that CDW quasiparticles are with large $\Gamma$ and eventually $\Gamma$ may exceed $m$ so that the quasi-particle picture breaks down \cite{Yan2020}. As a result, the term $\Gamma^2$ in the denominator of $\chi_{_{\rm CDW}}$ is large and responsible for the decrease in the observed RIXS intensity. The relaxation rate $\Gamma$ also defines the cut-off energy $\omega_{\rm c} \equiv {2\Gamma}$; if $\omega \gg \omega_{\rm c}$, the RIXS intensity is negligible because the term $i2\Gamma\omega$ is relatively small and ${\rm Im}\chi_{_{\rm CDW}}$ is insignificant.  

\begin{figure*}[ht!]
\centering
\includegraphics[width=2.1 \columnwidth]{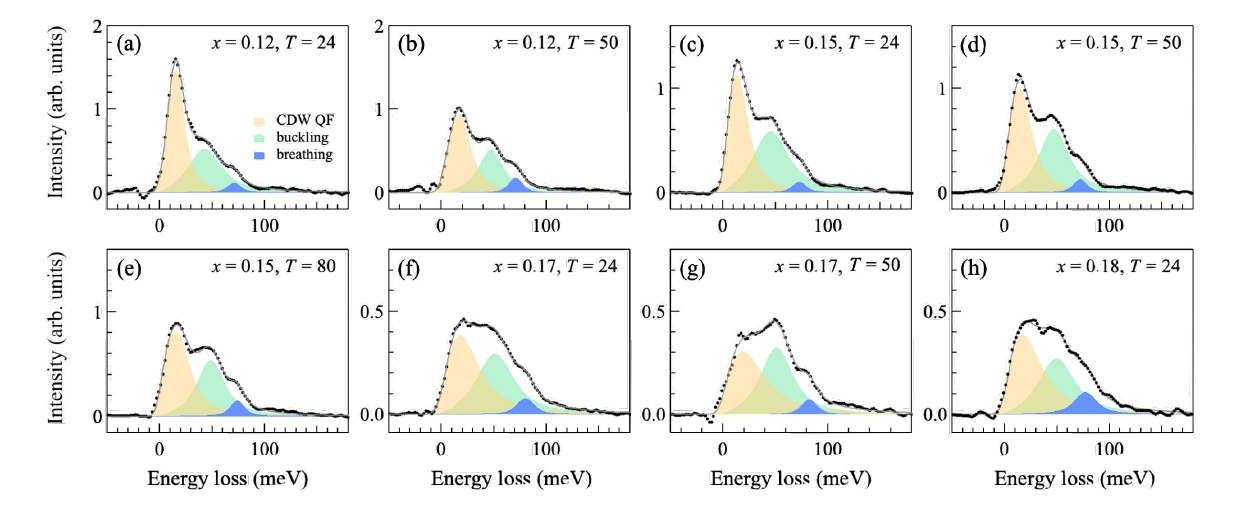}
\caption{Analysis of the RIXS spectral profile of CDW fluctuations in La$_{2-x}$Sr$_x$CuO$_4$ for ${q}_{\|}$ integrated from 0.23 to 0.24 at various temperatures and dopings. Selected RIXS spectra and spectral components from curve fitting are plotted in subplots: (a)-(b) for $x = 0.12$ at $T = 24~\&~50$~K;  (c)-(e) for  $x = 0.15$ at $T = 24, 50, \&~80$~K; (f)-(g) for  $x = 0.17$ at $T = 24~\&~50$~K; (h) for  $x = 0.18$ at $T = 24$~K. The plotted RIXS spectra are after the subtraction of elastic scattering. The curve-fitting scheme is the same as that of Fig. 3 except for the acoustic component, which was replaced by a component of CDW fluctuation. The dynamical structure factor due to CDW fluctuations is related to the charge susceptibility $\chi({\bf q}, {\omega})$ by $S({\bf q},\omega)=S_{0}(1-{\rm e}^{-\beta\omega})^{-1} {\rm Im} \chi({\bf q}, {\omega})$, in which $S_{0}$ is a proportion constant and $\beta = 1/k_{_B}T$, with $k_{_B}$ denoting the Boltzmann constant. Note that the reduced Planck constant $\hbar$ is set to 1 in the expression of $S({\bf q},\omega)$. 
The raw RIXS data of all doping levels and temperatures are plotted in Figs. S8 of the Supplementary Information.}\label{Delta_Gamma}
\end{figure*}

\begin{figure*}[ht!]
\centering
\includegraphics[width=2.1 \columnwidth]{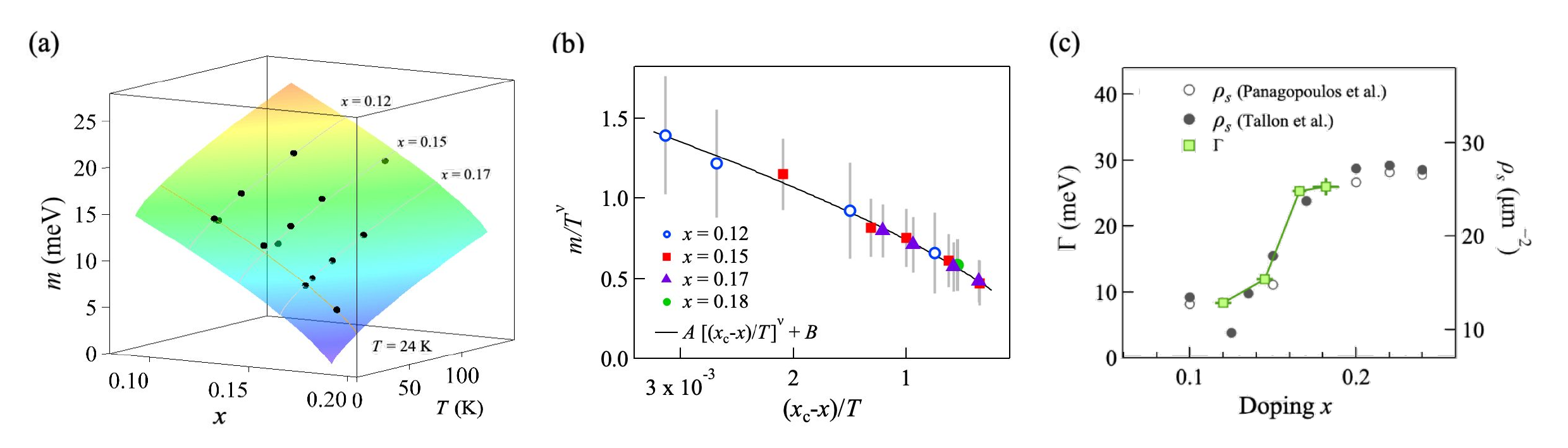}
\caption{Scaling of the inverse correlation length of the CDW fluctuations and the evolution of relaxation rate $\Gamma$ in La$_{2-x}$Sr$_x$CuO$_4$. (a), Doping and temperature dependence of the characteristic energy $m$. Black circles depict the deduced $m$ for various temperatures at $x = 0.12$, $0.15$, $0.17$, and $0.18$. They were fitted to a power-law form: $m  =  A(x_{\rm c}-x)^{\nu}+BT^{\nu}$ with a fitting scheme described in the Supplementary Information. The fitted coefficients were: $A = 81.68~{\pm}~16.2$, $x_{\rm c} = 0.195~{\pm}~0.007$, $\nu = 0.74~{\pm}~0.08$, and $B = 0.25~{\pm}~0.11$, defining the energy of $m$ shown by the color surface. Solid lines plot power-law curves of $(x_{\rm c}-x)^\nu$ for $T = 24$~K and $T^\nu$ for $x = 0.12$, 0.15, and 0.17. (b), Plot of $m/T^{\nu}$ versus $(x_{\rm c}-x)/T$ for various doping levels $x$ and temperatures $T$.  All data points, with minimal variations, collapse onto a curve that takes the power-law form of $Ay^{\nu}+B$, where $y = (x_{\rm c}-x)/T.$ The gray vertical bars represent the error bars of $m/T^{\nu}$, where the individual uncertainties of $m$ and $\nu$ were derived from the fitted errors and then combined through error propagation.  
(c), Doping-dependent relaxation rate compared with the doping dependency of superfluid density $\rho_s$, reproduced from Refs.~\onlinecite{Panagopoulos1999} (Panagopoulos et al.) and \onlinecite{Tallon2003} (Tallon et al.). The uncertainties of $\Gamma$ were derived from the fitted errors. $\rho_s$ is expressed in terms of $1/\lambda_{ab}^{2}$, where $\lambda_{ab}$ is the in-plane magnetic penetration depth. The solid line serves as a guide for the eye.} \label{scaling}
\end{figure*}

To find the critical exponent $\nu$ in the experiment, we analyzed the RIXS spectra for $q=0$ in terms of the dynamical structure factor $S({\bf q}_{\|}, \omega)$ derived from a charge susceptibility $\chi_{_{\rm CDW}}({\bf q}_{\|}, {\omega})$ taking the form given by Eq.~(1). For RIXS derived from CDW fluctuations, spectral characteristics including energy position, widths in $\omega$ and $q$, and relative intensity dictate the values of parameters $m$ and $\Gamma$ of our phenomenological analysis. Figures \ref{Delta_Gamma}(a)-\ref{Delta_Gamma}(h) present RIXS spectra of doping $x=0.15$ with curve fitting analysis at various temperatures, and those of $x=0.12,~0.17,~\&~0.18$ at $T = 24$~K. From these, we obtained doping- and temperature-dependent characteristic energy $m$ as indicated by black circles shown in Fig.~\ref{scaling}(a). We then simultaneously fitted  $m$  to the power-law  expression $m  =  A(x_{\rm c}-x)^{\nu}+BT^{\nu}$ for various dopings and  temperatures. The fitted critical exponent $\nu$ was $0.74 \pm 0.08$. Other fitted coefficients are given in the caption of Fig.~\ref{scaling}(a). 
To further verify the fitting of the exponent $\nu$, we resorted to the scaling analysis based on collapsing data from various dopings and temperatures according to Eq.~(2) with $\alpha=1$. Figure~\ref{scaling}(b) plots $m/T^{\nu}$ versus $|x-x_c|/T$ and shows that, with minimal variations, all data points collapse onto a single curve that takes the power-law form of $F(y)$ with $\nu = 0.74$, although $m(x,T)$ varies for different doping levels $x$. The successful collapse of the data for the inverse correlation lengths demonstrates the quantum critical scaling, which provides key evidence for the existence of a QCP in LSCO \cite{sachdev2011}. 
Note that the scaling behavior further indicates that the data points are captured in the critical regime of the quantum critical point. The critical regime does not need to be extremely close to the critical point \cite{chakravarty1989}. The large error bars shown in Fig.~\ref{scaling}(b) reflect the effects of fluctuations in the critical regime. Furthermore,  
the error of the exponent comes from the resolution of the measurement and uncertainty due to fluctuations in fitting to the proposed form of susceptibility $\chi_{_{\rm CDW}}$.

We further checked if the observed is reasonable. First, the CDW order in cuprates resides in a layered structure so that the effective dimension for QCP is D = 2 + 1, similar to the classical CDW transition in NbSe$_3$, in which the CDW order can be characterized by the three-dimensional O(2) model \cite{Moudden1990}. The observed value, $\nu=0.74 \pm 0.08$, satisfies $\nu \geqslant 2/D$; hence, according to the Harris criterion \cite{Harris1974}, the critical scaling is not affected by disorder and the QCP is governed by a clean model. The exponent value $\nu \sim 0.74$ is %clearly
not a mean-field value,  
consistent with our LSCO samples being already in the critical regime. Remarkably, the observed value of $\nu$ is close to that of the clean O($N$) model with $N \geq 2$ (for $N=2$, the $\epsilon$-expansion gives $\nu \simeq 0.67$ \cite{Shalaby2020}), indicating that orders other than the CDW order also participate.  Furthermore, the central value of the obtained $\nu$ is close to that of the O(4) model in which $\nu \simeq 0.74$ (from the $\epsilon$-expansion), rather than that of O(2) model \cite{Shalaby2020}. This suggests that the symmetry of the QCP is enlarged from O(2) symmetry to O(4) symmetry.  What would be the extra components in the enlarged symmetry?

To answer this question, we note that orders involved  in cuprates are generally complicated. In particular, it is known that the CDW order couples to many degrees of freedom in quite complicated ways including octahedral tilts, structural disorder, SDW order, the striple order, and superconductivity. However,  LSCO does not
undergo the low-temperature tetragonal phase transition and is an ideal platform for studying the CDW order \cite{WenNatComm2019}. The effect due to structure disorder can be ruled out by the Harris criterion shown above. Therefore, the possible candidates for the extra components are narrowed down to either charge- or spin-related orders. Experimentally, there is evidence of magnetic fluctuations \cite{zhu2023spin} and charge fluctuations \cite{Du2020} near the QCP. The remaining candidates for the extra components include spin density waves (SDW),  PDW, SC, stripe, and nematic order.
To further identify the extra components, we first note that the O $K$-edge of RIXS of cuprates is not sensitive to spin order. The contribution of orders with magnetic origin, if it exists, must be indirect through the spin-charge coupling. Furthermore, just like the phase transition, orders that contribute to the exponent $\nu$ for the quantum critical point must be static. Fluctuating orders that couple indirectly to the charge order do not contribute to $\nu$. However, it is known that SDW is not a static order around $x=x_c$ and $T=0$ \cite{WenNatComm2019} and even for the fluctuating stripe order that contains SDW, the ordering wavevector is not ${\bf Q}$. 
Secondly, we note that as shown in the Supplementary Information, the derivation that leads to Eq.~(1) can be generalized to O($N$) models so that Eq.~(1) is still applicable with $\nu$ characterizing the O($N$) model.  
Furthermore, in the O(4) symmetry, the four components of the order parameter are symmetric and play the same role. 
As a result, the above analysis implies that extra components are in the charge sector. 
Further hinted by the observation of PDW signatures shown in Fig. S9 of the Supplementary Information and those observed in La-based cuprates doped with Sr or Fe via resonant X-ray scattering \cite{Lee2023PDW}, %and in Bi2212 by STM using a superconducting tip \cite{Du2020}, 
we consider the PDW orders for the additional components in the enlarged symmetry.

Indeed, in the Ginzburg-Landau theory \cite{Agterberg2008, Xu2018}, there are two possible PDW orders that go with the CDW order and could be stablized as thermodynamic orders: $\Delta_{\nicefrac{\textbf{Q}}{2}}$ and $\Delta_{\textbf{Q}}$, where $\Delta_{\textbf{P}}$ represents the PDW order parameter that describes particle-particle pairing with a total momentum of {\textbf P}. These PDW orders interact with CDW and SC orders through the interactions: $I_1 \equiv \rho_{\textbf{Q}} \, \Delta_{\nicefrac{-\textbf{Q}}{2}} \Delta^{*}_{\nicefrac{\textbf{Q}}{2}} +h.c.$ and $I_2= \rho_{\textbf{Q}} \, \Delta^{*} \Delta_{-\textbf{Q}}+ h.c.$.  For $\Delta_{\nicefrac{\textbf{Q}}{2}}$, $I_1$ is present when $T$ is above $T_c$. We have checked 
its existence (see Fig.~S9 in the Supplementary Information), consistent with the results shown in Ref. \onlinecite{Lee2023PDW}.  For $T<T_c$, $I_1$ needs to combine with the superconducting order $\Delta$. The lowest non-trivial term would be $I_{1} \times |\Delta|^2$, which is a higher order term near the critical point. Hence it can be neglected. As a result, the extra components arise from $\Delta_{\textbf{Q}}$. In this case, as the superconducting order $\Delta$ is a fixed background order, $I_2$ is a bilinear term of $\rho_{\textbf{Q}}$ and $\Delta_{-\textbf{Q}}$, which, when combined with $|\rho_{\textbf{Q}}|^2$ and $|\Delta_{\textbf{Q}}|^2$, forms the quadratic terms of the combined order $(\rho_{\textbf{Q}}, \Delta_{\textbf{Q}})$ for the O(4) model. Furthermore, the momentum in $\rho_{\textbf{Q}}$ is also consistent with the momentum carried by the CDW order and with the experimental findings \cite{Du2020}. 
Therefore, we conclude that the observed QCP belongs to the universality class characterized by the O(4) symmetry with the order parameter being ($\rho_{\textbf{Q}}$,$\Delta_{\textbf{Q}}$).  
Note that the found O(4) symmetry reminisces us of the microscopic SO(4) symmetry in the Hubbard model at half filling \cite{Yang1989,Yang1990}, where the CDW order is also combined with the superconducting order to form the order parameter. The O(4) symmetry of the QCP may thus be rooted from the microscopic SO(4) symmetry.

As indicated above, a unique feature of the QCP in cuprate superconductors is that it is buried in the SC phase, resulting in all transitions involved occurring within the superconducting state. Hence one expects that quantum critical fluctuations are induced by intertwining of orders, mostly occurring among SC, CDW, and PDW orders. The effect of intertwining is reflected in Fig.~\ref{scaling}(c), which plots the doping-dependent relaxation rate $\Gamma$ obtained from curve fitting for $T = 24$~K. We found that $m$ reduces as the hole doping approaches the critical doping, while the relaxation rate $\Gamma$ increases.
This observation of CDW dynamics supports the QCP scenario and offers an explanation for the difficulty in detecting the QCP due to the enhanced scattering by the quantum fluctuations. In addition, the quasiparticle inverse lifetime versus doping is shown to exhibit similar trend as that of the superfluid density. This behavior can be obtained by using the phenomenological form of charge susceptibility derived in the Supplementary Information, where it is shown that $\xi^{-2}$ results from correlations of phase gradient of the CDW orders. By the coupling of the phase gradient of the CDW order to that of the SC order, it is easy to see that the self-energy (thus the inverse lifetime) of quasiparticle is proportional to the correlation of phase gradient of the SC order. Since the phase gradient of the SC order is proportional to the supercurrent, the inverse lifetime is proportional to supercurrent-supercurrent correlation which is then proportional to the superfluid density.

The model of the QCP in LSCO given by Eq.(1) fits our experimental data successfully. It differs from the model of the Ohmic dissipative QCP \cite{LeeNatPhys2021},  which has been  %speculated 
used to interpret the weakening and broadening of the observed CDW peak in cuprate superconductors. For an Ohmic dissipative QCP, due to disorders, the Ohmic dissipation occurs during the tunneling  between large clusters near the critical point. In this scenario, the weakening and broadening of the CDW peak is attributed to the dissipation of the order parameter through its quantum fluctuations and the generation of fermionic particle–hole excitations. However, unlike Eq.(1), the charge susceptibility of the Ohmic dissipative QCP takes the form $(m^2+c^{2}q^{2}+\gamma|\omega|)^{-1}$ \cite{Hoyos2007},
where the linear term, $\gamma|\omega|$, is generated by fermionic particle–hole excitations.  
This form of $\chi_{_{\rm CDW}}({\bf q}_{\|}, {\omega})$ 
exhibits a shape kink at $\omega = 0$ and thus doesn't fit our data.
On the contrary, in our model based on Eq.(1), the weakening and broadening of the observed CDW peak is attributed to the increase of $\Gamma$ such that it exceeds $m$ as the doping $x$ approaches $x_{\rm c}$, the QCP. This behavior is known as the breakdown of bosonic quasiparticle picture and is consistent with a similar observation in a Bose-Einstein condensate within the temperature domain \cite{Yan2020}. The breakdown of the CDW quasiparticle in cuprates results from quantum fluctuations near the QCP enhanced by the intertwining of SC, CDW, and PDW \cite{Larkin1965,Fulde1964,Agterberg2020,TuSciRep2019,Liu2023}. Our findings suggest %provide
a new perspective for comprehending the quantum phase transition in cuprates. The transition involves two phases: below the QCP, a state characterized by intertwined CDW, PDW, and SC; above the QCP, a pure $d$-wave SC state with maximum superfluid density.

\vspace{10mm}
\noindent\textbf{Acknowledgements}: 
The authors acknowledge Chung-Hou Chung, Yu-Te Hsu, and Gregory S. Boebinger for their insightful discussions. We also thank Ganesha Channagowdra for his assistance with the PDW measurements shown in the Supplementary Information. This work was supported in part by the National Science and Technology Council of Taiwan under Grant Nos. NSTC113-2112-M-007-033, NSTC113-2112-M-213-016, NSTC112-2112-M-007-031, NSTC112-2112-M-213-026-MY3 and NSTC110-2923-M-213-001. We also thank the support by the Japan Society for the Promotion of Science under Grant No. JP22K03535. AF acknowledges the support from the Yushan Fellow Program under the Ministry of Education (MOE) of Taiwan. CYM acknowledges support from the Center for Quantum Science and Technology within the framework of the Higher Education Sprout Project by the MOE of Taiwan.

\vspace{5mm}
\noindent\textbf{Author contributions}:
DJH coordinated the project. HYH, AS, JSS, JO, DJH, and CTC conducted the RIXS experiments. SK synthesized and characterized the LSCO samples.  HYH and DJH analyzed the RIXS data. CYM developed the phenomenological model of charge susceptibility. DJH, CYM, AF, HYH, and TKL wrote the manuscript with inputs from other authors.

\vspace{5mm}
\noindent\textbf{Competing interests}:
Authors declare that they have no competing interests.

\vspace{5mm}
\noindent\textbf{Data and materials availability}:
All raw data are plotted in the main text or the Supplementary Information. Their numeric values are available from the corresponding author upon a reasonable request.

%\bibliography{reference}


\end{document}


\baselineskip24pt


\maketitle 
\vspace{5mm}
\noindent {\bf This SI file includes:}\\
\indent Materials and methods\\
\indent Curve-fitting analysis of RIXS data\\
\indent Derivation of correlation function for CDW amplitude fluctuation\\
\indent References 1 to 6\\
\indent Figures S1 to S9
\newpage

\section{Materials and methods}

The La$_{2-x}$Sr$_x$CuO$_4$ single crystals with various doping levels $x$ were grown by the traveling-solvent floating zone method \cite{Komiya2002, Komiya2005, Ono2007}. After growth, the crystals were annealed to remove oxygen defects. The precise values of doping concentration were determined from an inductively-coupled-plasma atomic-emission spectrometric analysis. For the samples of nominal doping $x= 0.12,~0.15,~0.17,$~and~$0.18$, their precise dopings were $0.12\pm0.005$, $0.145\pm0.005$, $0.166\pm0.004$ and $0.182\pm0.008$, respectively. The $T_{_{\rm C}}$ of the $x=0.15$ sample was 37.5~K. 

We conducted O $K$-edge resonant inelastic X-ray scattering (RIXS) measurements using the AGM-AGS spectrometer of beamline 41A at Taiwan Photon Source of the National Synchrotron Radiation Research Center, Taiwan \cite{Singh2021}. This AGM-AGS beamline was constructed based on the energy compensation principle of grating dispersion. The best energy resolution at incident photon energy of 530~eV was 16 meV. The scattering angle was fixed to 150$^\circ$. %The angle between the incident X-ray and the ${\rm ab}$ plane of the sample is $\theta$.  
The wave vectors of incident and scattered X-rays are ${\bf k}_{\rm i}$ and ${\bf k}_{\rm f}$, respectively. The momentum transfer is $\mathbf{q}= {\bf k}_{\rm i} - {\bf k}_{\rm f}$, and its projection onto the ${\rm ab}$ plane is  $\mathbf{q}_\|$. The $\rm a$-axis and $\rm c$-axis lay in the horizontal scattering plane while the $\rm b$-axis was perpendicular to the scattering plane.

Prior to XAS and RIXS measurements, the crystallographic axes were aligned with hard X-ray diffraction using a special holder with tilting adjustment, and then the sample was cleaved in air to have a $(001)$ surface. The resonant conditions were achieved by tuning the energy of the incident X-ray to the O $\textit{K}$-edge absorption peak around 528.5 eV, which arises from the resonance to the mobile $p$ holes hybridized with the Cu $3d_{x^{2} - y^{2}}$ orbital to form a spin singlet termed Zhang-Rice singlet. RIXS spectra were recorded using $\sigma$ polarized incident X-rays of which the polarization was perpendicular to the scattering plane. The total energy resolution, characterized by the full width half maximum of elastic scattering, was between 16 and 25 meV, due to the use of a bendable grating in the spectrometer. For every RIXS spectrum, we also measured reference spectra from a reference sample—carbon tape—before and after the real measurement to determine the zero-energy position and the instrument’s energy resolution, which were used in subsequent curve fitting to ensure accurate data analysis. Despite these variations, the energy resolution ${\Delta}E$ of our RIXS measurements remained better than 25 meV.

\section{Curve-fitting analysis of RIXS data}

Figures S2-S5 present the raw data of RIXS for each momentum scan measured at $T = 24$~K for nominal doping levels $x = 0.12, 0.15, 0.17$, and $0.18$. RIXS raw data of $x=0.15$ measured at $T = 38$, $50$, $80$, and $140$~K were presented in our previous publication \cite{HYHuang2021}. %[12]: Huang et al. Phys. Rev. X 11, 041038 (2021).

The curve-fitting analysis of all RIXS spectra  were conducted through a non-linear least square scheme using a scientific data analysis software Igor Pro.  We used an iterative method which applies the Levenberg-Marquardt algorithm to minimize the chi-square $\chi^2$, defined as the sum of the squared differences between the measured data points $(y_{i})$ and the corresponding calculated values $(f_{i})$ from a model function, normalized by the uncertainties $(\sigma_{i})$ of the data points $(i)$, i.e., 
\begin{equation}
\chi^{2} = \sum_{i}^{N}\left(\frac{f_{i}-y_{i}}{\sigma_i}\right)^{2},
\end{equation}
where $N$ is the total number of data points.  The model functions used in the curve-fitting analyses of Figs.~3(a)-3(d) in the main text include a Voigt function for elastic scattering and three spectral functions of damped harmonic oscillator (DHO) for phonon excitations. The spectral function $f_{_{\rm DHO}}(\omega)$ of energy loss $\omega$ arising from DHO is 
\begin{equation} \label{DHO}
f_{_{\rm DHO}}({\omega})= f^{0}_{_{\rm DHO}}\frac{\gamma\omega}{(\omega^{2}-\omega_{0}^{2})^2+4 \omega^{2}\gamma^{2}},
\end{equation}
where $\omega_0$ and $\gamma$ are the bare phonon energy and the damping constant, respectively. $f^{0}_{_{\rm DHO}}$ is a proportion constant.  Additionally, to reflect the finite spectral width contributed by the energy resolution of the spectrometer, each DHO component is convoluted with a Gaussian profile with a full width defined by the full width of the Voigt profile fitted to elastic scattering.

In the curve fitting of Figs.~4(a)-4(h) in the main text and Fig.~\ref{fitting_QF}, the DHO function centered around 15 meV was replaced by the dynamical structure factor $S({\bf q},\omega)$ due to charge-density wave (CDW) fluctuations of wave vector ${\bf Q}$ related to the charge susceptibility 
\begin{equation} \label{chi}
\chi_{_{\rm CDW}}({\bf q}, {\omega})=  \frac{1}{m^2+c^{2}q^{2}-(\omega+i\Gamma)^2},
\end{equation}
where $m$ denotes the characteristic energy of the dynamical CDW, $q$ is the momentum magnitude $q = |{\bf q}_{\|}- {\bf Q}|$, $c$ is a parameter characterizing the speed of CDW excitation, and $\Gamma$ characterizes the inverse lifetime for the excitation. That is, $S({\bf q},\omega)=S_{0}(1-{\rm e}^{-\beta\omega})^{-1} {\rm Im} \chi({\bf q}, {\omega})$. $\beta = 1/k_{_B}T$, with $k_{_B}$ denoting the Boltzmann constant. Note that the reduced Planck constant $\hbar$ is set to 1 in the expression of the dynamical structure. The fitted spectral energies and widths of phonon components of ${\bf q}_{\|}$ away from ${\bf Q}$ imposed constraints on those of the curve-fitting analysis of RIXS derived from CDW quantum fluctuations. First, we analyzed the phonon energy and width (damping factor $\gamma$) for $q_{\|}$ between 0.1 and 0.22. Figure \ref{rixs_phonon} plots the dispersion of phonon energy $\omega_0$ for the breathing and buckling modes at various dopings and $T = 24$~K, consistent with previous results of Cu $L$-edge RIXS \cite{LinPRL2020}. Figures~3(a)-3(d) and Fig. \ref{q_0p22} show the fitted spectra for $q_{\|} = 0.12$ and $0.22$, respectively. We used the fitted $\omega_0$ and $\gamma$ of $q_{\|} = 0.22$ for the breathing and buckling phonons in the fitting analysing of CDW fluctuations.

To find the critical exponent $\nu$, we simultaneously fitted  both doping- and  temperature-dependent $m$  to a power-law  expression: $ m  =  A(x_{\rm c}-x)^{\nu}+BT^{\nu}$. 
Figure 5(a) in the main text plots the deduced values of $m$ for various temperatures and dopings. The coefficients simultaneously fitted were: $A = 81.68~{\pm}~16.2$, $x_{\rm c} = 0.195~{\pm}~0.007$, $\nu = 0.74~{\pm}~0.08$, and $B = 0.25~{\pm}~0.11$. The error bars of the fitted coefficients are given based on a confidence interval 95\%. 

\section{Derivation of correlation function for CDW amplitude fluctuation}

We start from the partition function that describes the fluctuation of the order parameter for CDW in the 2D CuO$_2$ plane 
\begin{equation}
Z=  {\int D \rho({\bf r},t)}   {\rm e}^{\frac{i}{\hbar} S[ \rho({\bf r},t)]} , \label{S1}
\end{equation}
where $S$ is the action %and is 
given by the integration of the Lagrangian density $L$ as $S=\int \int  dt d^2 {\bf r}L$, and $\int D \rho({\bf r},t)$ refers to the functional integral.  
To derive $L$, one first notes that for the charge density $\rho({\bf r},t)$ to propagate as a wave,  it must satisfy a wave equation at the linear level:
\begin{equation} 
(\partial^2_t - c^2 \nabla^2) \rho ({\bf r},t)=0, \label{CDWwave}
\end{equation}
where $c$ is the speed of charge density wave. Using this fact and combining it with the Ginzburg-Landau theory, in the lowest order term, $L$ can be written as
\begin{equation} 
L= \alpha(T-T_c) |\rho|^2 + \rho^* (\partial^2_t - c^2 \nabla^2) \rho +\frac{g}{2} |\rho|^4, \label{S2}
\end{equation}
where $\rho$ is the order parameter of the CDW, $T_c =T_c(x)$ is a function of doping $x$, the first and the third terms in together form the Ginzburg-Landau Hamiltonian for uniform $\rho$, and the second term is the generalization of the usual kinetic energy term and gives rise to the wave dynamics of CDW shown in Eq.(\ref{CDWwave}). Note that after the integration by parts, the second term becomes $-(\partial_t - c\nabla)\rho^* \cdot (\partial_t - c\nabla) \rho$. For the uniform $\rho$, the second term drops and one finds the mean field solution $\rho_0= \sqrt{\alpha[T_c (x)-T]/g} e^{i\theta_0 }$ with $\theta_0$ being a constant phase. Hence $\alpha[T_c (x)-T]= g |\rho_0|^2$. By expressing $\rho$ as $\rho_0 + |\delta \rho|e^{i \theta}$ in $L$ and replacing the integration $\int D \rho({\bf r},t)$ by $\int D \delta \rho({\bf r},t)$, one finds that the lowest-order Lagrangian density $L$ for the amplitude $|\delta \rho|$ is given by 
\begin{equation} 
L=[ g |\rho_0|^2 -  (\partial_t - c \nabla )\theta \cdot (\partial_t -c \nabla )\theta ] |\delta \rho|^2 + |\rho| (\partial^2_t - c^2 \nabla^2) |\delta \rho|. \label{S3}
\end{equation}
From Eq.~(\ref{S3}), one finds that the correlation function of amplitude $ \langle |\delta \rho| ({\bf r} ,t) |\delta \rho| ({\bf r}',t')\rangle$ is given by
\begin{equation} 
\langle |\delta \rho| ({\bf r} ,t) | \delta \rho| ({\bf r}',t')\rangle = \frac{1}{g |\rho_0|^2 + (\partial^2_t - c^2 \nabla^2)  -  (\partial_t - c \nabla )\theta \cdot (\partial_t -c \nabla )\theta}. 
\end{equation}
In the lowest-order approximation, one may replace $ (\partial_t - c \nabla )\theta \cdot (\partial_t -c \nabla )\theta$ by its average. Hence in the Fourier space, one obtains
\begin{equation} 
\langle |\delta \rho| ({\bf q} ,\omega) |\delta \rho| (-{\bf q},-\omega)\rangle = \frac{1}{g |\rho_0|^2  -  \langle (\partial_t - c \nabla )\theta \cdot (\partial_t -c \nabla )\theta \rangle +c^2 q^2 -\omega^2}.  \label{S5}
\end{equation}
Here after the Wick rotation, $(\partial_t - c \nabla )\theta$ becomes the 3D gradient of $\theta$. For finite temperatures, one expects $\langle \delta \rho^*({\bf r},t) \delta \rho({\bf r}',t') \rangle \propto  e^{-|\bf{R}-\bf{R}'|/\xi}$ (Here, ${\bf R}$ is a vector that combines ${\bf r}$ and $t$, ${\bf R}=({\bf r}, ct)$).  By taking 3D gradient ($\nabla$) on the above equation and expressing $\delta \rho$ as $|\delta \rho|{\rm e}^{i \theta}$, one finds $\nabla \theta \sim 1/\xi$. Therefore, one expects that  $ -  \langle (\partial_t - c \nabla )\theta \cdot (\partial_t -c \nabla )\theta \rangle $ in Eq.~(\ref{S5}) is proportional to $\xi^{-2}$. Furthermore, by going back to the variable ${\bf r}$ and $t$, one expects $\langle \delta \rho^*({\bf r},t) \delta \rho({\bf r}',t') \rangle \propto  e^{-|{\bf r}-{\bf r}'|/\xi_x-c|t-t'|/\xi_t}$ at doping $x$ close $x_c$ and temperatures region in close to zero temperature. 
Here $\xi_x \sim |x-x_c|^{-\nu}$ is the correlation lenth projected at zero temperature for $x \sim x_c$ and  $\xi_T   \sim  T^{-\nu}$ is the correlation length projected along the temperature axis near zero temperature.
Consider the correlation function on the circle of $|\bf{R}-\bf{R}'|$ centered at $({\bf r}', ct')$ with ${\bf R - R'}=\left({\bf r}-{\bf r}', c(t-t')\right)$,  generally $|{\bf r}-{\bf r}'| = |\bf{R}-\bf{R}'| |\cos \theta|$ and $ c|t-t'| = |\bf{R}-\bf{R}'| |\sin \theta|$, we find $\langle \delta \rho^*({\bf r},t) \delta \rho({\bf r}',t')\rangle\propto  e^{-|{\bf r}-{\bf r}'|/\xi_x-c|t-t'|/\xi_T}$$= e^{-(|\cos \theta|/\xi_x+|\sin \theta |/\xi_T)|\bf{R}-\bf{R}'|}$. By taking the average over the angle $\theta$, i.e., $|{\bf r}-{\bf r}'|/\xi_x+c|t-t'|/\xi_T$ is replaced by $\langle |{\bf r}-{\bf r}'|/\xi_x+c|t-t'|/\xi_T \rangle$ with $\langle \cdots \rangle$ representing the average over angle, we find that 
$e^{-\langle |\cos \theta|/\xi_x+|\sin \theta |/\xi_T) \rangle | \bf{R}-\bf{R}'|  } \sim e^{- \frac{2}{\pi} (\xi^{-1}_x+\xi^{-1}_T) |\bf{R}-\bf{R}'|}$. Since one expects $\langle \delta \rho^*({\bf r},t) \delta \rho({\bf r}',t') \rangle \propto  e^{-|\bf{R}-\bf{R}'|/\xi}$, we conclude that $\xi^{-1} = \frac{2}{\pi} (\xi^{-1}_x+\xi^{-1}_T)$.
%because $\partial_t \theta$ is corrected to $ \sim 1/\xi_T$, one expects $ \xi^{-2} = \xi_x^{-2}+ \xi_T^{-2}$.
% as the direction of correlation length can be either in the time (temperature) direction $\xi_T$ or in the spatial direction $\xi_x$ (which will be doping dependent) , it implies $ \xi^{-2} = \xi_T^{-2} +\xi_x^{-2}$.   e^{-|\bf{R}-\bf{R}'|/\xi-|t-t'|/\xi_T}
Finally, Eq.~(1) in the text can be recovered by appropriately rescaling $\rho_0$ so that $g$ is one. Note that results of the above derivation are consistent with Ref. 20 but are also applicable to the case when the relevant order parameter $O$ for the quantum critical point is $n$-component. In this situation, one simply replaces $\delta \rho$ by $\delta O$ so that  $\langle \delta O^*({\bf r},t) \delta O({\bf r}',t')\propto  e^{-|{\bf r}-{\bf r}'|/\xi_x-c|t-t'|/\xi_T}$.  Furthermore, if $n$-component of $O$ includes $\rho$ and the system is isotropic in components (which is the case for O(4) symmetry), then one would get the same result $\langle \delta \rho^*({\bf r},t) \delta \rho({\bf r}',t')\propto  e^{-|{\bf r}-{\bf r}'|/\xi_x-c|t-t'|/\xi_T}$.

\newpage
\bibliographystyle{Science_fixed_title.bst}
\bibliography{reference}

\newpage

\begin{figure}[t!]
\centering
\includegraphics[width=1 \columnwidth]{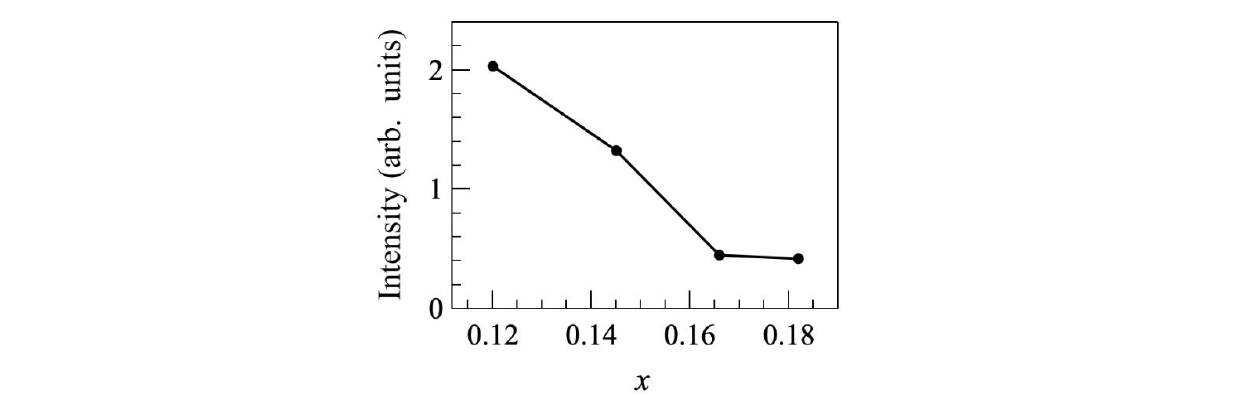}
\caption{The measured RIXS intensity of CDW fluctuation is not enhanced as the QCP is approached.}
\label{fig_S1}
\end{figure}

\begin{figure}[t!]
\centering
\includegraphics[width=1 \columnwidth]{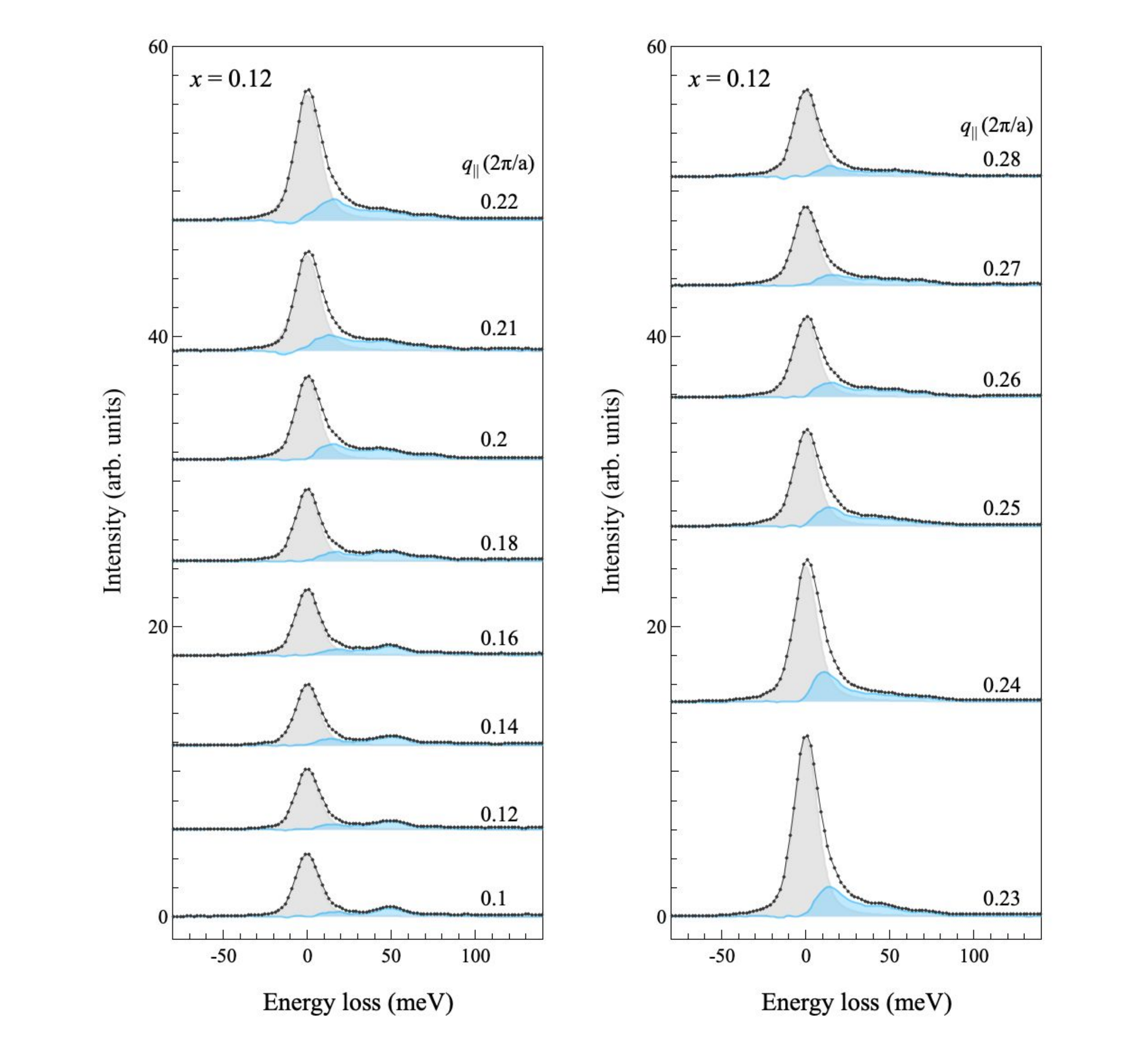}
\caption{Momentum-dependent RIXS spectra of La$_{2-x}$Sr$_x$CuO$_4$ for $x = 0.12$ measured at $24$~K. The black circles are raw data and the solid line serves as a guide for the eye. The spectral component of elastic scattering is highlighted in gray and the RIXS spectra after the subtraction of elastic scattering is highlighted in light blue.}
\label{rixs_x12}
\end{figure}

\begin{figure}[t!]
\centering
\includegraphics[width=1 \columnwidth]{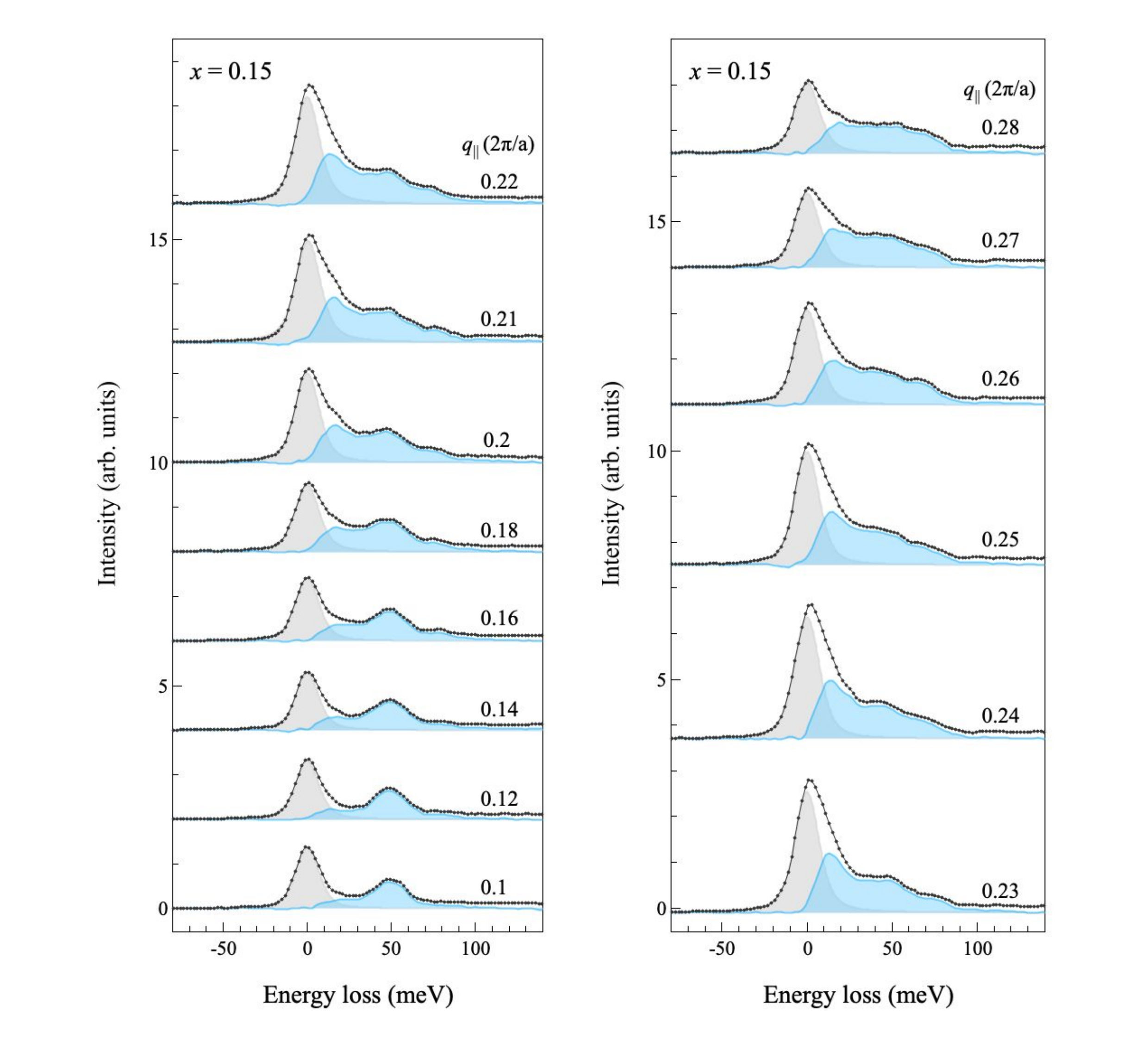}
\caption{Momentum-dependent RIXS spectra of La$_{2-x}$Sr$_x$CuO$_4$ for $x = 0.15$ measured at $24$~K. The black circles are raw data and the solid line serves as a guide for the eye. The spectral component of elastic scattering is highlighted in gray and the RIXS spectra after the subtraction of elastic scattering is highlighted in light blue.}
\label{rixs_x15}
\end{figure}

\begin{figure}[t!]
\centering
\includegraphics[width=1 \columnwidth]{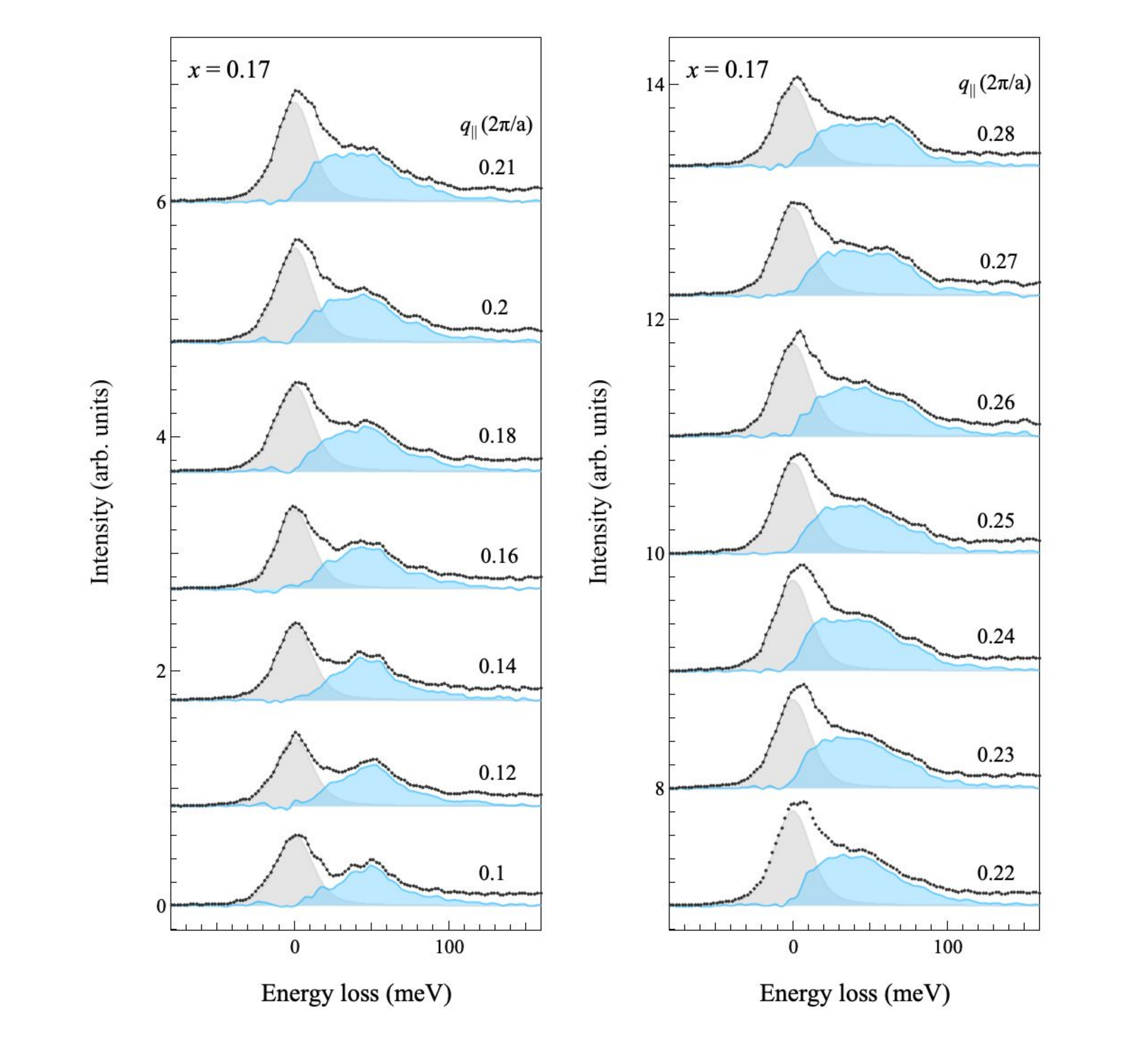}
\caption{Momentum-dependent RIXS spectra of La$_{2-x}$Sr$_x$CuO$_4$ for $x = 0.17$ measured at $24$~K. The black circles are raw data and the solid line serves as a guide for the eye. The spectral component of elastic scattering is highlighted in gray and the RIXS spectra after the subtraction of elastic scattering is highlighted in light blue.}
\label{rixs_x17}
\end{figure}

\begin{figure}[t!]
\centering
\includegraphics[width=1 \columnwidth]{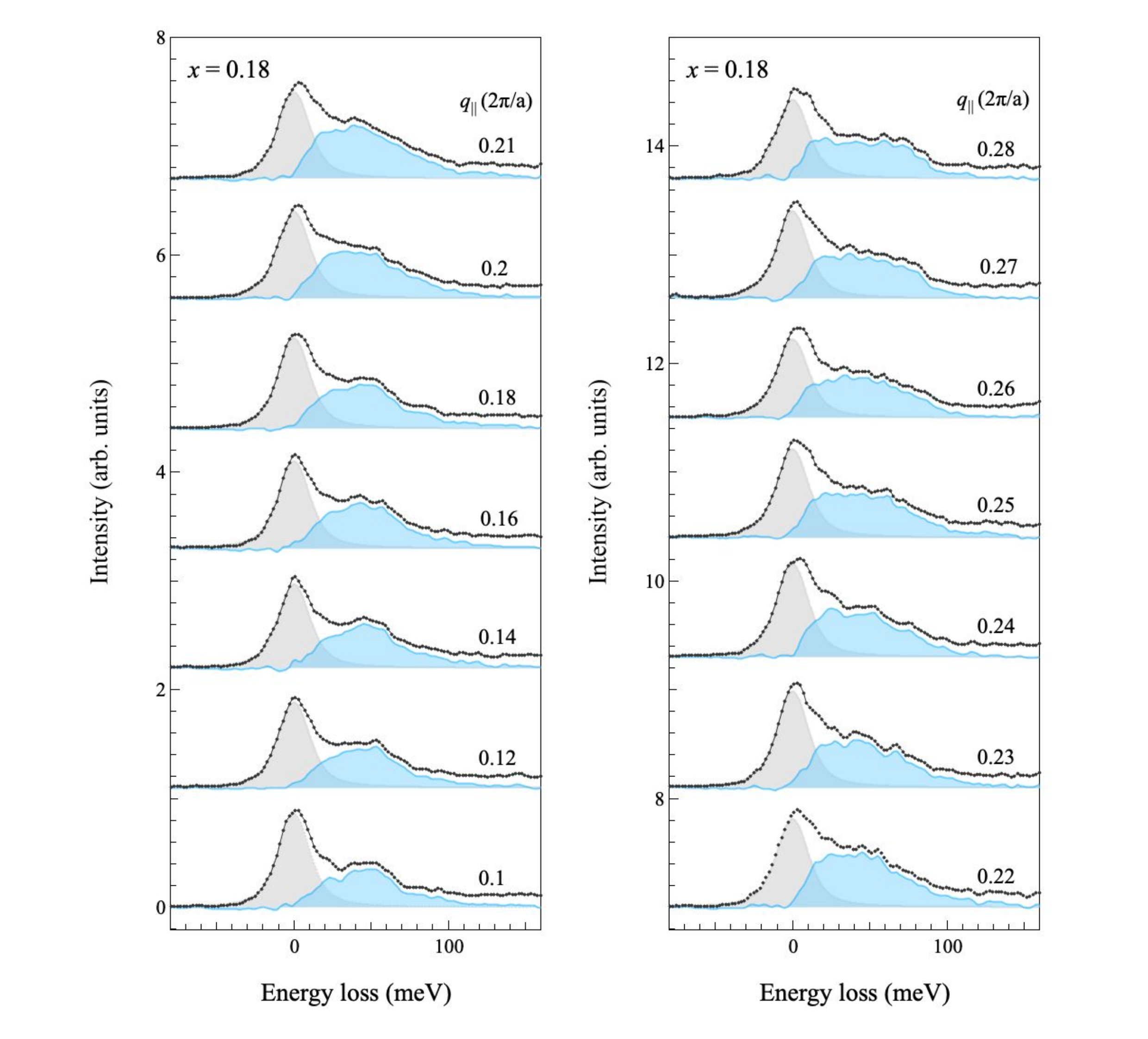}
\caption{Momentum-dependent RIXS spectra of La$_{2-x}$Sr$_x$CuO$_4$ for $x = 0.18$ measured at $24$~K. The black circles are raw data and the solid line serves as a guide for the eye. The spectral component of elastic scattering is highlighted in gray and the RIXS spectra after the subtraction of elastic scattering is highlighted in light blue.}
\label{rixs_x19}
\end{figure}

\begin{figure}[t!]
\centering
\includegraphics[width=1 \columnwidth]{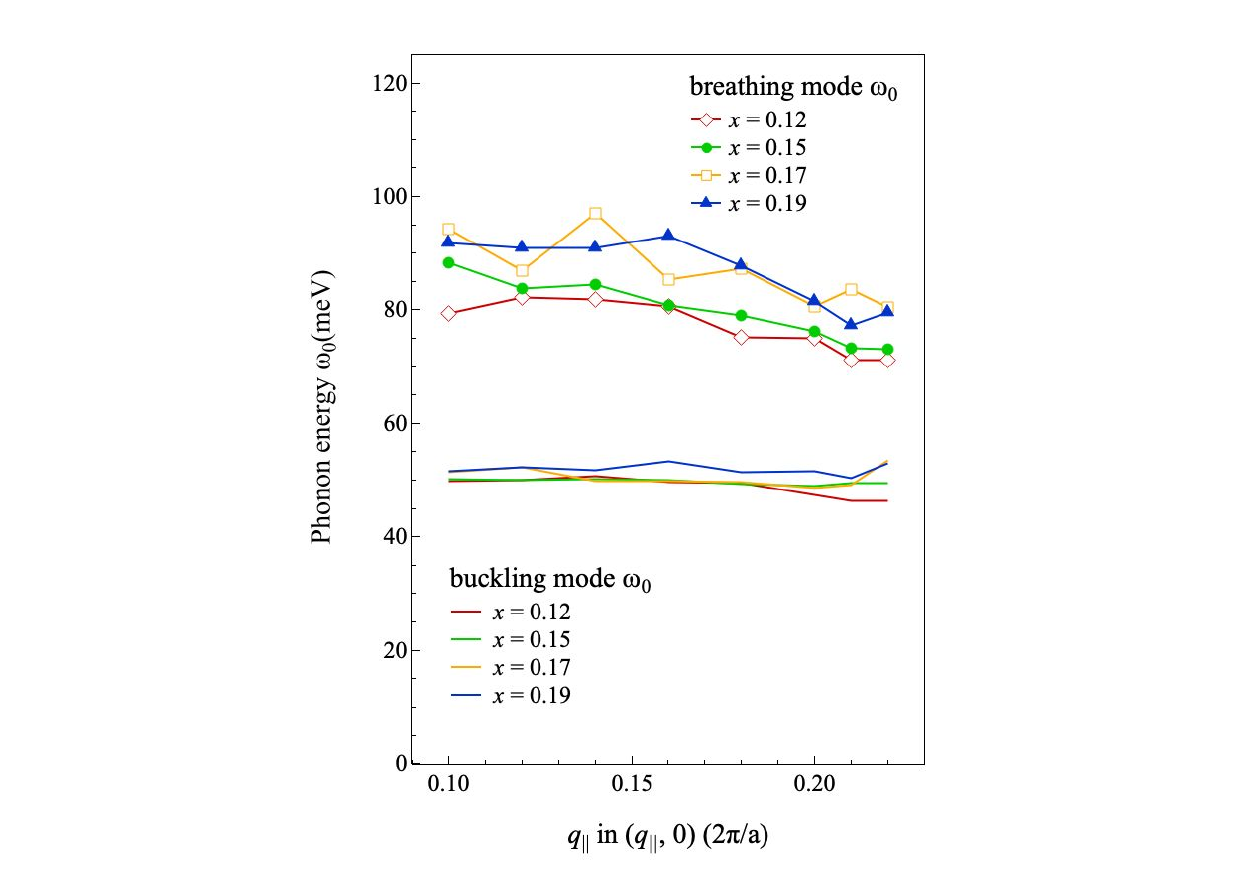}
\caption{Momentum-dependent phonon energies from RIXS of La$_{2-x}$Sr$_x$CuO$_4$ for various doping $x$ measured at 24~K.}
\label{rixs_phonon}
\end{figure}

\begin{figure}[t!]
\centering
\includegraphics[width=1 \columnwidth]{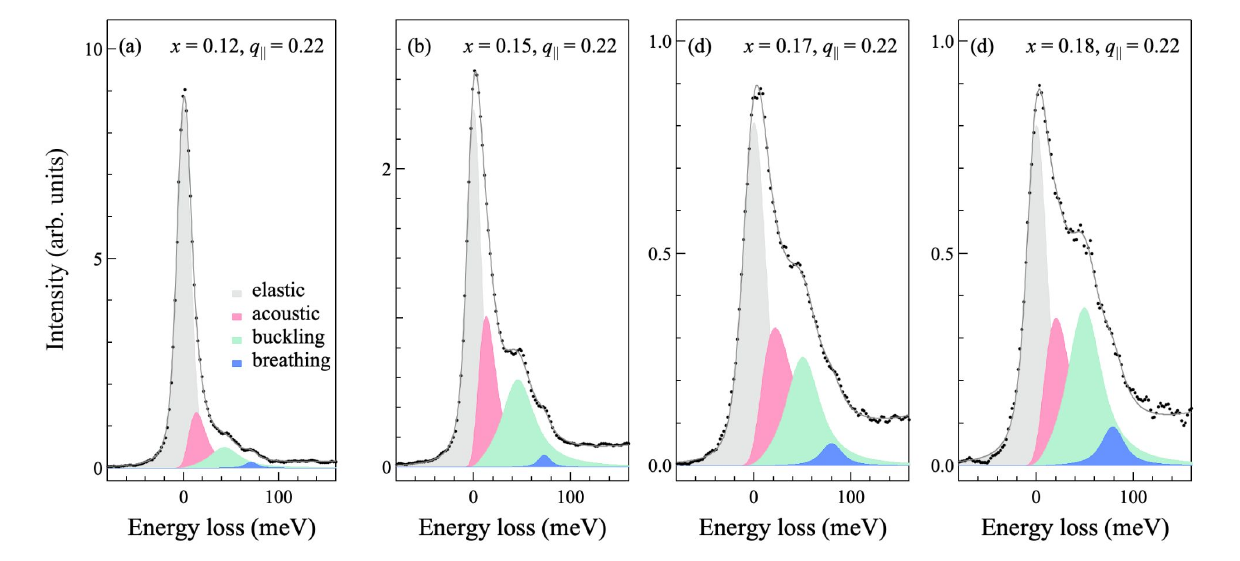}
\caption{Curve-fitting analysis of $q_{\|} = 0.22$ RIXS spectra for $x = 0.12$, $0.15$, $0.17$, and $0.18$ measured at $24$~K. The RIXS spectra were fitted to four components with a linear background: elastic scattering, acoustic phonons, a mix of buckling phonon and apical oxygen phonon, and half-breathing (bond-stretching) phonons, shaded in gray, pink, green, and blue, respectively. Each phonon component was fitted to a spectral function of damped harmonic oscillator. RIXS data are plotted as black circles, and the fitted curve is given as a gray line. The fitted values of  $\omega_0$ ($\gamma$) of the breathing phonons were $71.07, 73.06, 80.47,$ and $79.5$ meV ($5, 5, 9$, and $11$ meV) for $x = 0.12$, $0.15$, $0.17$, and $0.18$, respectively. The corresponding values of the buckling phonons were $\omega_{0} = 46.4, 49.5, 53.5,$ and $52.9$~meV, and  $\gamma = 18.6, 19.7, 18.8$, $19.6$ meV, respectively.}   
\label{q_0p22}
\end{figure}

\begin{figure}[t!]
\centering
\includegraphics[width=0.95 \columnwidth]{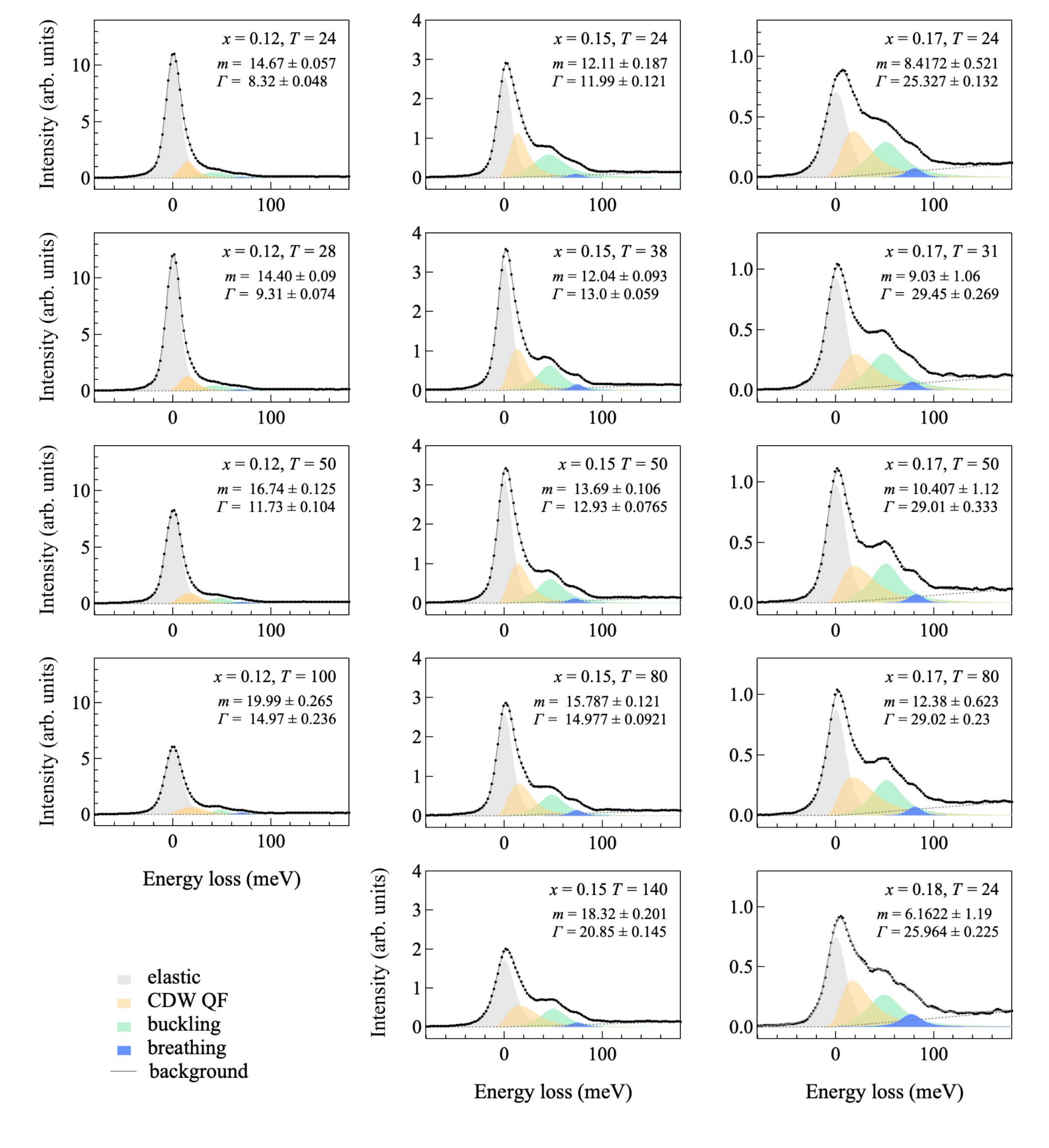}
\caption{Curve-fitting analysis of RIXS spectra measured at various temperatures (in units of K) with momentum $q_{\|}$ integrated from $0.23$ to $0.24$ for $x = 0.12$, $0.15$, $0.17$, and $0.18$. The RIXS spectra were fitted to four components with a linear background, including elastic scattering, CDW fluctuations, a mix of buckling phonon and apical oxygen phonon, and half-breathing (bond-stretching) phonons, shaded in gray, orange, green, and blue, respectively. The dynamical structure factor due to CDW fluctuations is related to the charge susceptibility $\chi({\bf q}, {\omega})$ by $S({\bf q},\omega)=S_{0}(1-e^{-\beta\omega})^{-1} {\rm Im} \chi({\bf q}, {\omega})$. The error bars of fitted coefficients are given based on a 50\% confidence interval.} %Curve-fitting analysis of RIXS spectra for $x = 0.12$, $0.15$, $0.17$, and $0.18$ measured at various temperatures (in units of K). The RIXS momentum integration range was from $0.23$ to $0.24$. RIXS spectra were fitted into four components with a linear background: elastic scattering, CDW fluctuation, the mix of buckling phonon and apical oxygen phonon, and half-breathing (bond stretching) phonons, represented in gray, organge, green, and blue shades, respectively. The dynamical structure factor due to CDW fluctuations is related to the charge susceptibility $\chi({\bf q}, {\omega})$ by $S({\bf q},\omega)=S_{0}(1-e^{-\beta\omega})^{-1} {\rm Im} \chi({\bf q}, {\omega})$.}
\label{fitting_QF}
\end{figure}

\begin{figure}[t!]
\centering
\includegraphics[width=0.9\columnwidth]{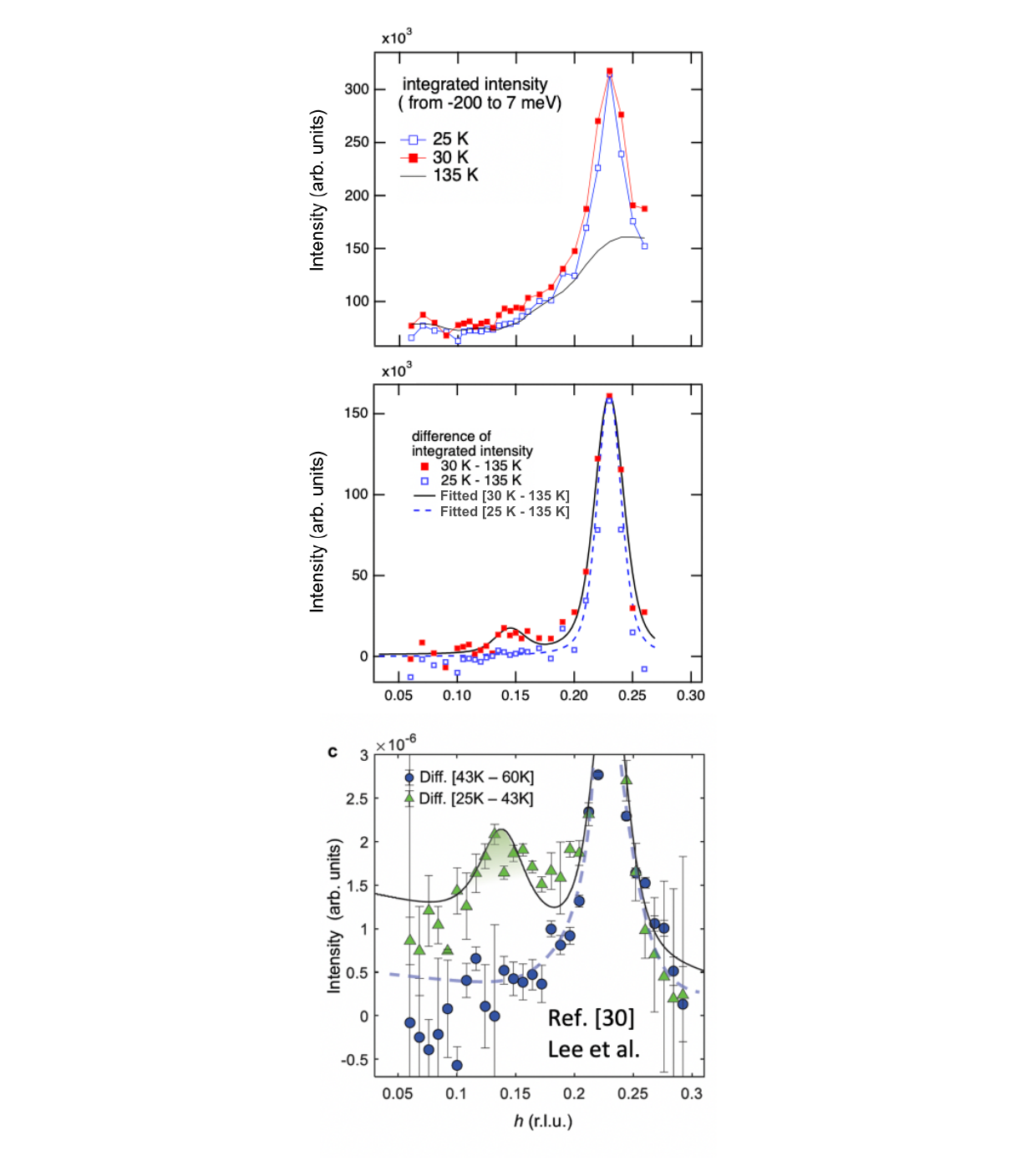}
\caption{RIXS evidence of PDW for La$_{2-x}$Sr$_x$CuO$_4$ (LSCO) with $x = 0.12$. Energy-integrated RIXS intensity plotted against in-plane momentum $q_{|}$ for energies ranging from $-200$~meV to 7~meV at temperatures of 25, 35, and 135~K.
Middle: Energy-integrated RIXS at 25 and 35 K, as shown in the top panel, after subtraction of the reference RIXS at 135~K. Bottom: Difference spectra reproduced from Fig.~3c of Lee et al., arXiv:2310.19907 (2023), which plots the differences between resonant soft X-ray scattering (RSXS) measured at 25, 43, and 60~K in Fe-doped LSCO, La$_{1.87}$Sr$_{0.13}$Fe$_{0.01}$CuO$_4$ with a $T_{_C}$ below 6~K.}
\end{figure}

%\begin{figure}[t!]
%\centering
%\includegraphics[width=1 \columnwidth]{s9_fitting_LSCO15.pdf}
%\caption{Curve-fitting analysis of RIXS spectra for $x = 0.15$ measured at 38 K, 50 K, 80 K and 140 K. The RIXS momentum integration range was from $0.23$ to $0.24$. The curve-fitting and color schemes are the same as those of Fig. \ref{fitting_24K}. }
%\label{fitting_0p15}
%\end{figure}

%\begin{figure}[t!]
%\centering
%\includegraphics[width=1 \columnwidth]{s9_fitting_LSC.pdf}
%\caption{Curve-fitting analysis of RIXS spectra for $x = 0.15$ measured at 38 K, 50 K, 80 K and 140 K. The RIXS momentum integration range was from $0.23$ to $0.24$. The curve-fitting and color schemes are the same as those of Fig. \ref{fitting_24K}. }
%\label{fitting_0p15}
%\end{figure}

%\begin{figure}[t!]
%\centering
%\includegraphics[width=1 \columnwidth]{Fig_SM_m_vs_x_T_0410.pdf}
%\caption{Simultaneous fitting of $m$ to a power form: $ m  =  A(x_{\rm c}-x)^{\nu}+BT^{\nu}$.}
%\label{m_xT}
%\end{figure}